\documentclass[preprint2]{aastex}

\shorttitle{Nonthermal radiation of young SNRs} \shortauthors{V.N.
Zirakashvili et al.}

\begin{document}


\title{NONTHERMAL RADIATION OF YOUNG SUPERNOVA REMNANTS: THE CASE OF CAS A}


\author{V.N.Zirakashvili}
\affil{Pushkov Institute for Terrestrial Magnetism, Ionosphere and Radiowave
Propagation, 142190, Troitsk, Moscow Region, Russia}
\affil{Max-Planck-Institut f\"{u}r\ Kernphysik, Saupfercheckweg 1, 69117
Heidelberg, Germany}
%\email{zirak@mpimail.mpi-hd.mpg.de}
\author{F.A.Aharonian}
\affil{Dublin Institute for Advanced Studies, 31 Fitzwilliam Place, Dublin 2, Ireland}
\affil{Max-Planck-Institut f\"{u}r\ Kernphysik,  Saupfercheckweg 1, 69117
Heidelberg, Germany}
\author{R.Yang}
\affil{Max-Planck-Institut f\"{u}r\ Kernphysik,  Saupfercheckweg 1, 69117
Heidelberg, Germany}
\author{E.O$\tilde{\mathrm{n}}$a-Wilhelmi}
\affil{Max-Planck-Institut f\"{u}r\ Kernphysik,  Saupfercheckweg 1, 69117
Heidelberg, Germany}
\author{R.J.Tuffs}
\affil{Max-Planck-Institut f\"{u}r\ Kernphysik,  Saupfercheckweg 1, 69117
Heidelberg, Germany}



\begin{abstract}
The  processes responsible for the broad-band radiation  of  the young
supernova remnant Cas~A are explored  using a  new code which
is  designed for a detailed  treatment of the
diffusive shock acceleration of particles in nonlinear regime.  The model is based on spherically
symmetric hydrodynamic equations complemented with transport
equations for relativistic particles.  Electrons, protons and the oxygen ions
accelerated by forward and reverse
shocks are  included  in the numerical calculations.
We show that the available multi-wavelength  observations in the
radio, X-ray and gamma-ray bands can be best explained by invoking
particle acceleration  by both forward and reversed shocks.
Although the TeV gamma-ray observations can be interpreted by
interactions of  both accelerated electrons and protons/ions, the
measurements by Fermi LAT at energies below 1~GeV give a tentative
preference to the hadronic origin of gamma-rays. Then, the
acceleration efficiency in this source, despite the previous
claims, should be very high;  25 \% of the  explosion energy (or
approximately $3\cdot 10^{50}$~erg) should already be  converted
to cosmic rays, mainly by the forward shock. At the same time, the
model calculations do not provide extension of the maximum energy
of accelerated protons beyond  100~TeV.  In this model, the
acceleration of electrons  is dominated  by the reverse shock; the
required $10^{48}$~erg can be achieved under the assumption  that
the injection of electrons  (positrons) is supported  by the
radioactive decay of  $^{44}$Ti.

\end{abstract}


%% subject header.  Objects should be in the appropriate "individual"
%% headers (e.g. quasars: individual, stars: individual, etc.) with the
%% additional provision that the total number of headers, including each
%% individual object, not exceed six.  The \objectname{} macro, and its
%% alias \object{}, is used to mark each object.  The macro takes the object
%% name as its primary argument.  This name will appear in the paper
%% and serve as the link's anchor in the electronic edition if the name
%% is recognized by the data centers.  The macro also takes an optional
%% argument in parentheses in cases where the data center identification
%% differs from what is to be printed in the paper.


%% In the first two sections, notice the use of the natbib \citealpp
%% and \citealpt commands to identify citations.  The citations are

\section{Introduction}

The  mechanism of diffusive shock acceleration (DSA) of relativistic particles
 (Krymsky \citealp{krymsky77}, Axford et al. \citealp{axford77}, Bell
\citealp{bell78}, Blandford \& Ostriker \citealp{blandford78})  is generally accepted
as the  most likely  paradigm  for production of
galactic cosmic rays (CR) in supernova remnants (SNRs). Over the last 30 years
a significant progress has been achieved in the development of theoretical models and
understanding  of  the  basic features of DSA   (see e.g. Malkov \& Drury
\citealp{malkov01}, Schure et al. \citealp{schure12} for a review).  On the other hand, the recent detailed studies
of spectral and morphological features of young SNRs, first of all in the
X-ray and very-high-energy (or TeV)  gamma-ray band, provide excellent observational
material for  development of detailed  numerical models of acceleration and radiation
of relativistic electrons and protons in young SNRs.  These observations generally confirm
in general terms the  predictions of DSA. In particular  the synchrotron X-radiation
observed from  several  young
SNRs implies an existence of multi-TeV electrons which is naturally explained by DSA.
The detection of TeV gamma-rays from SNRs, like Cas~A, RX J1713.7-3946, Vela Jr.,
RCW~86 (see Hinton \& Hofmann \citealp{hinton09}; Rieger et al. \citealp{rieger13}
for a recent review) give a more direct and unambiguous
information about the effective acceleration of particles, electrons and/or protons, in SNRs to energies
exceeding 100 TeV.

In this paper we  conduct detailed study of acceleration of electrons
and protons with an emphasis on the  spectral and morphological
features of high energy radiation produced by these particles in the
young supernova remnant Cas~A.  For that purpose we use a new numerical
code of nonlinear diffusive shock acceleration developed by one of
us in collaboration with V. Ptuskin (Zirakashvili \& Ptuskin
\citealp{zirakashvili12}). This model can be considered as a
natural development of existing numerical codes (see e.g. Berezhko
et al.\citealp{berezhko94}, Kang et al \citealp{kang06}),  with
new additional elements  which despite their  strong impact on the
overall picture of acceleration in general, and on the properties
of  high energy radiation of SNRs in particular, have been ignored
in the past.   Namely, in our  treatment, the solution of
spherically symmetric hydrodynamic equations is combined with the transport
and acceleration of  relativistic particles  by the {\it forward}
shock and {\it reverse}  shocks  (FS and RS, respectively).  The nonlinear response of
energetic particles via their pressure gradient results in a
self-regulation of acceleration efficiency.   The   detailed calculations  of radio,
X-ray, gamma-ray emission components conducted within  a
self-consistent treatment of the particle acceleration by both forward
and reverse shocks  should allow a direct comparison of the observed
spectral and morphological features with model predictions.
The inclusion of the radiation components related to the reverse
shock seems to be  a rather obligatory condition, at least for the specific case of Cas~A.
In this regard we note that  the parameters  that characterize the reverse
shock can be significantly different compared to  the
parameters of  the forward shock. Therefore, the properties of
radiation components from the reverse and forward shocks are also expected to be
significantly  different. In particular,  the magnetic field
in the reverse shock can be  very
small  which would dramatically increase the contribution of the IC
component compared to the hadronic ($\pi^0$-decay) component  of
gamma-rays.   Because of the stronger
magnetic field,  in the forward shock  just opposite
relation is expected between the contributions of electrons  and  protons
to production of high energy gamma-rays.

An interesting feature of Cas~A  is  the  non-negligible  contribution  of the
decay of  radioactive elements in the ejecta to the  production of suprathermal
electrons and positrons as a potentially important ``injection material''. The acceleration of these particles by
the reverse shock (Zirakashvili \& Aharonian \citealp{zirakashvili11}, Ellison et al. \citealp{ellison90})
can result in a rather high energy content of leptonic
component in SNRs. In particular this can
explain the high electron to proton  ratio found in Cas~A (Atoyan et al. \citealp{atoyan00}).

The results and conclusion of this study  have a rather general
character and can be applied to different SNRs.  In this paper
the model is used to interpret the multi-wavelength properties of
Cas A - one of the youngest supernova remnants in our Galaxy. The
high quality  X-ray  images and energy spectra, as well as the
coverage of gamma-ray  observations from low to very high
energies, provide adequate observational material to conduct
detailed theoretical studies of acceleration and radiation
processes in this unique source.

The paper is organized as follows. In Section 2 we briefly summarize multi-wavelength observations of
Cas A. The short description of the model is  given in Section  3. The results of
modeling of the broad-band emission
are presented  and discussed in Sections  4, 5, 6 and 7. Finally,
in Sections 8 and 9 we discuss and summarize the obtained results.

\section{Observational properties of Cas A}

The supernova explosion  related to  Cas~A is likely to be linked to the event
observed by Flamsteed in 1680 (Ashworth \citealp{ashworth80}). Recently Krause et al.
\cite{krause08} reported the detection  of its light echo with a spectrum
similar to the spectra of IIb  supernova 1993J.  Therefore one may assume  that
SNR Cas A  was  produced in a IIb type supernova explosion.
Generally,  the  progenitors of such  explosions are
Red Super Giants (RSG) which already have lost  their  hydrogen envelope via
a powerful stellar wind (Chevalier \citealp{chevalier05}).

Cas A  has been  extensively observed at radio wavelengths (Bell et al.
\citealp{bell75}; Tuffs \citealp{tuffs86}; Braun et al.
\citealp{braun87}; Anderson et al. \citealp{anderson91}; Kassim et
al. \citealp{kassim95}; etc.). The radio spectrum  is
close to power-law $J(\nu )\sim \nu ^{-\alpha }$ with a spectral  index
$\alpha =0.77$ (Baars et al. \citealp{baars77}). The main fraction  of
radio-emission comes from the bright radio-ring with an angular
radius close to $100''$ and from the faint outer radio-plateau
with an angular radius $150''$.  For a distance  to the source $D=3.4$ kpc
(Reed et al. \citealp{reed95}), the corresponding spatial radii
are 1.7 pc and 2.5 pc, respectively.

In addition to the large-scale structures,  several hundreds of very compact and
bright radio-knots with steeper spectra (Anderson et al. \citealp{anderson91})
are present in the radio-shell.
It is believed that the bright radio-ring is related to  the
reverse shock propagating into the supernova ejecta.
A large number of  fast-moving knots (FMK) are observed also
in optics (e.g. Fesen et al. \citealp{fesen88}) are attributed to densest clumps
of ejecta which are not strongly decelerated after the supernova
explosion.
This hypothesis agrees  with X-ray observations which show that
the X-ray line emitting shell roughly
coincides with the bright radio ring. The X-ray emitting plasma is rich in
 O, Si, Ar, Ca and Fe dominated by the contribution from oxygen
(Fabian et al. \citealp{fabian80};
Markert et al. \citealp{markert83};
Vink et al. \citealp{vink96}, Hughes et al. \citealp{hughes00}; Willingale et al.
\citealp{willingale02,willingale03}; Hwang \& Laming \citealp{hwang03}; Laming \& Hwang \citealp{laming03}).

Besides the line X-ray emission,   Cas A shows a hard X-ray continuum extending up to 100~keV with a photon index  $\sim 3$  as measured by {\it BeppoSAX} (Vink et al.
\citealp{vink01}), {\it INTEGRAL} (Renaud et al.
\citealp{renaud06})  and  {\it Suzaku} observations
(Maeda et al. \citealp{maeda09})  satellites. Unfortunately,  because of limited
angular resolution of these instruments,  the production region(s) of the component of radiation cannot be localized. At low energies,
thin and faint  nonthermal X-ray filamentary structures
have been  found  at the periphery of the
remnant close to the boundary of the radio plateau  (Gotthelf et
al. \citealp{gotthelf01}). These filaments correspond to the
position of the forward shock propagating in the circumstellar medium.
The proper motion of these filaments has been  measured by the
{\it Chandra}  observations allowing the estimate  of the
forward shock velocity of about 4900 km s$^{-1}$
(Patnaude \& Fesen \citealp{patnaude09}).

On the other hand, the inner shell in the radio and X-ray images can be naturally attributed to the reverse shock
propagating in the supernova ejecta, assuming  that the electrons are accelerated also at the
reverse shock  (Uchiyama \& Aharonian \citealp{uchiyama08}, Helder \& Vink \citealp{helder08}).
Generally, the reverse shock is
not treated  as an efficient accelerator, because the magnetic field of ejecta might be very weak
due to the large  expansion factor of the exploded star. However, similar to the case of the forward shock,
the  magnetic field can be  significantly amplified  also at the reverse shock
(see Ellison et al. \citealp{ellison05}), for example,
in the course of the nonresonant streaming instability  as suggested  by Bell  \cite{bell04}.
This implies that we should expect gamma-rays from both forward and reverse shocks
the contributions of which however  cannot be separated by current  gamma-ray telescopes both in GeV and TeV energy bands.

Very high energy gamma-ray emission from Cas~A  has been discovered by the
HEGRA system of atmospheric Cherenkov telescopes
(Aharonian et al. \citealp{aharonian01}), and later  confirmed by the MAGIC
(Albert et al. \citealp{albert07}) and  VERITAS (Acciari et al. \citealp{acciari10}) collaborations.
The fluxes
published  by three groups are  in a reasonably good agreement with each other, and indicate on a
not-very-hard and not-very-soft  energy spectrum with a  photon index $\gamma =2.4-2.6$ and a maximal energy of 
 detected photons 5 TeV.

At low energies, a  weak gamma-rays signal  from Cas~A  has  been discovered
 after the first  year of observations  with  the {\it Fermi} Gamma-ray Space Telescope (Abdo et al. \citealp{abdo10}).
While the flux and the energy spectrum
reported  at energies above 0.5~GeV  are quite informative for constraining
certain model parameters,  they are not sufficient for a robust conclusions concerning, in particular, the
origin of  radiation. In this regard, the derivation of the
energy spectrum at energies below 1 GeV  seems to be  crucial
 for identification of the radiation mechanism(s). In particular,
 a  sharp decline  of the spectrum below 1~GeV  would indicate the dominance of
 hadronic interactions, at least  in this energy band.
Motivated by the  importance of such spectral measurements,
we attempted to reanalyze  the
{\it Fermi} LAT data  based on  much larger statistics accumulated over
 4~yr observations,  and using  the recent
 {\it Fermi} LAT software  package which allows the extension of the analysis down to 100~MeV.
The approach used in this paper  for analysis of the {\it Fermi} LAT data is described in 
\cite{yang12}.

We have selected the  events with
energy between 60~MeV and 100\,GeV and applied the usual filters recommended by the
{\it Fermi} LAT collaboration (removing the intervals when the rocking angle of the LAT was
greater than 52~deg or when parts of the region-of-interest (ROI) were
observed at zenith angles  larger than 100~deg). To derive the energy spectrum we applied the
maximum likelihood method  in 13 independent energy bins from 100~MeV to
100~GeV;  the Galactic and isotropic diffuse emission was modeled using the tools
{\it gal-2years7v6-v0} and  {\it iso-p7v6source}. During the broad-band fit, all
sources in the second Fermi catalog (2FGL) in  the field of view have been  included in
likelihood model. Points with
significance of more than 3$\sigma$ are shown in Fig. 6, 7,8 and 14. It should be
noted that below 200~MeV, the spectrum determination suffers heavily from the
background subtraction due to the uncertainties in the model of the Galactic
diffuse emission. Applying different models generated by  the GALPROP simulator,   we
obtained a 15 \%  larger flux that can be account by systematic errors.
The obtained spectral  energy distribution (SED)  shows  a significant decrease of the flux
below  1 GeV  which is expected in the SED of gamma-rays from decays of neutral
pions  produced at interactions of accelerated protons and nuclei with the ambient gas.

\section{Nonlinear model of diffusive shock acceleration}

Throughout this
paper we use the approach which lies on a
nonlinear  treatment of particle acceleration by strong shock waves
proposed  by  Zirakashvili \& Ptuskin \cite{zirakashvili12}.

Non-steady hydrodynamical equations for the gas density  $\rho (r,t)$, gas velocity $u(r,t)$, gas pressure
$P_g(r,t)$, and the equation for the quasi-isotropic CR momentum distribution
 $N(r,t,p)$ in the spherically symmetrical  case are solved numerically with the use of
a finite-difference method. Thermal protons are injected at the forward shock with a radius $r=R_f(t)$.
Since it is believed that the ejecta material in Cas A consists mainly on heavy elements,
 the oxygen ions are injected at the reverse shock
with a radius $r=R_b(t)$\footnote{Throughout the paper we use
the  subscripts "f"  and "b"  to  the parameters characterizing the forward and reverse
(backward) shocks, respectively.}.

The injection efficiency that is the fraction of particles
injected into DSA at the reverse and forward shocks $\eta _b$ and
$\eta _f$ is taken to be independent of  time, and the injection
momenta are $p_{f}=2m(\dot{R}_f-u(R_f+0,t))$,
$p_{b}=2M(u(R_b-0,t)-\dot{R}_b)$. Here $m$ and $M$ are the mass of
the proton and oxygen ion respectively. We normalize the momentum
distribution of oxygen ions $N_i$ to
 the number density of nucleons. Thus the number density of oxygen ions with atomic mass $A=16$
is given by $n_i=4\pi \int p^2dpN_i/A$.

%A  high injection efficiency of the order of $\eta =0.01$ may
%result in significant shock modification already at early epochs of the SNR expansion while the
%thermal sub-shock compression ratio is close to 2.5. This situation is probably realized at the
%quasiparallel shocks where the amount of injected particles is high enough and the self-regulation of
%acceleration efficiency by the pressure of accelerated particles will result in the sub-shock compression
%ratio close to 2.5. The corresponding non-relativistic momentum distribution $N(p)\propto p^{-5}$ has the
% equal amounts of energy density in the every decade of momentum.

To obtain the sub-shock compression ratio close to 2.5-2.7,
 for  simulations of cosmic ray modified shocks we adopt
 the injection efficiency $\eta _b=0.01$.
This is in agreement with
calculations  of  quasi-parallel  collisionless shocks
(Zirakashvili \citealp{zirakashvili07}) and is supported
by  radio-observations of young extragalactic  Ib/c SNRs
(Chevalier \& Fransson \citealp{chevalier06}). The amplified magnetic field is rather
large in these SNRs  and the synchrotron
radio-emission is produced by electrons with energies below 1 GeV. The observed radio-flux
$J(\nu )\propto \nu ^{-1}$ has a power-law
dependence on the frequency $\nu $. Note that since the radiative
losses  of radio electrons are negligible, this
corresponds to a $E^{-3}$ type energy  spectrum of accelerated electrons.
Such a spectrum can be  formed at the sub-shock with a compression ratio 2.5.

In older remnants, the magnetic field strength drops and  the energy spectra of
GeV electrons   become flatter
due to the higher compression ratio "seen" by these particles.
This change of the electron  energy distribution  is reflected in the spectrum of radio emission

In SNR shells  the ion injection may be suppressed at oblique shocks
(see V\"olk et al. \cite{voelk03} for the discussion of this
topic). This can be realized in
Cas A  where the forward shock propagates in the
stellar wind  with an azimuthal magnetic field.
That is why below we consider two extreme cases
with $\eta _f=0.01$ (hadronic model) and $\eta _f=10^{-7}$ (leptonic model).

Usually the pressure of energetic electrons is neglected, i.e. the
electrons are treated as test particles. However, recently it has been proposed
that gamma-rays from the radioactive decay of $^{56}$Ni
may produce a large number of energetic electrons via Compton scattering in
the circumstellar medium. In addition the decays  of $^{44}$Ti
can directly supply  energetic electrons and positrons in supernova ejecta
 (see Zirakashvili \& Aharonian \citealp{zirakashvili12} for details). These energetic particles are
further accelerated by the forward and reverse shocks. Under these
conditions,  the reverse shock may be modified by the pressure of
electrons (positrons). For this reason  in calculations we include the  contribution of
relativistic electrons to the overall pressure .

The evolution of
electron distribution is described by an equation similar to transport equation for protons,
but  with additional terms describing synchrotron, Coulomb, bremsstrahlung  and IC losses.
In the scenario of electron injection through the  decays of radioactive nuclei
$^{56}$Ni and $^{44}$Ti, the amount of positrons and electrons injected  at the forward and
reverse shock is  described by the injection efficiencies $\eta
^e_f$ and $\eta ^e_b$ respectively. For the forward shock the
injection efficiency is determined by the probability of
appearance of an energetic electron due to the Compton
scattering of gamma-rays from
 $^{56}$Co decay and is given by (Zirakashvili \& Aharonian
\citealp{zirakashvili11})

\begin{equation}
\eta ^e_f=1.2\cdot 10^{-7}\frac {M_{\rm Ni}}{M_{\odot }}r^{-2}_{\rm pc}.
\end{equation}
Here  $M_{\rm Ni}$ is the mass of $^{56}$Ni just after the
explosion and $r_{\rm pc}$ is the distance from the
center of the remnant expressed in parsecs. $^{56}$Co is produced via decay of $^{56}$Ni.

For the reverse shock the injection efficiency of electrons and positrons
 produced directly from
the decay of $^{44}$Ti
can be written as (Zirakashvili \& Aharonian \citealp{zirakashvili11})

\begin{equation}
\eta ^e_b=1.94\frac {M_{Ti}}{44M_{ej}}\left( 1-\exp
\left( -\frac {t\ln 2}{t_{1/2}}\right) \right) .
\end{equation}
Here $M_{ej}$ is the mass of ejecta, $M_{\rm Ti}$ is the mass of $^{44}$Ti just after the explosion,
and $t_{1/2}=63$ yr is the
half-time of $^{44}$Ti. It was taken into account that  one electron and
 0.94 positrons appear per a  decay of $^{44}$Ti.

CR diffusion is determined by magnetic inhomogeneities. Strong
streaming of accelerated particles changes medium properties in
the shock vicinity. In particular, the CR streaming instability in young SNRs
 results in a high level of magnetohydrodynamical (MHD) turbulence  (Bell \citealp{bell78})
and even in  the amplification of magnetic field  (Bell \citealp{bell04}). Due
to this effect protons can be accelerated to energies beyond the
the so-called  Lagage and Cesarsky  upper limit,
$E \sim  100$~TeV (Lagage \& Cesarsky \citealp{lagage83}).

The magnetic field amplified upstream  the shock is enhanced further via compression in the
 shock transition region. It can even play a dynamical role downstream of the shock. We take
 magnetic pressure and magnetic energy flux into account downstream of the shock. This is
 a new element in comparison with the approach of Zirakashvili \& Ptuskin \cite{zirakashvili12}
 where the magnetic field spatial distribution was prescribed.
The magnetic field was transported in the downstream region as the gas with 
adiabatic index $\gamma _m$.
Its impact  on the shock dynamics was also taken into account via Hugoniot conditions.
Upstream of the forward shock where dynamical effects of the magnetic fields are small,
the coordinate dependence of the magnetic field $B$ can be  prescribed as :
\begin{equation}
B(r)=\sqrt{4\pi \rho _0}\frac {V_f}{M^f_A}\left( \frac {\rho (r)}{\rho _0}\right) ^{\gamma _m/2},
\end{equation}
Here $\rho _0$ and $\rho (r)$ are the undisturbed gas density at the shock position
and the density of the
medium where the shock propagates respectively, $V_f$ is the speed of the forward shock.
The parameter
  $M^f_A$ is similar to the Alfv\'en Much number of the shock and
determines the value of the amplified magnetic field strength
far upstream of the shock. In the shock transition region the magnetic
field strength is increased by a  factor of $\sigma ^{\gamma _m/2}$,  where
$\sigma $ is the shock compression ratio. An  expression  similar
to Eq. (3) is used  also in the upstream region of the reverse shock.

Below we
use the adiabatic index of isotropic random magnetic field $\gamma _m=4/3$. For this value of the
adiabatic index
 the magnetic pressure $P_m=B^2/24\pi $ is three times smaller
 than the  magnetic energy density ($=B^2/8\pi $).
%{\bf reference?}

%The
%magnetic energy downstream of the shock is estimated to be several percents of
%the dynamical pressure $\rho _0\dot{R}^2$ as was derived from the
%width of X-ray filaments in young SNRs (V\"olk et al.
%\citealp{voelk05}). The characteristic range of this parameter is
%$M^f_A\sim 10\div 70$. For example, the energy of the magnetic
%field of about 3.5 percent  of the dynamical pressure  (V\"olk et
%al. \citealp{voelk05}) and characteristic compression ratio of a
%modified shock
% $\sigma =6$ correspond to  $M^f_A\approx 23$.

CR advective velocity may differ from the gas velocity by  the value of the radial
component of the Alfv\'en velocity
$V_{Ar}=V_A/\sqrt{3}$ calculated in the isotropic random magnetic field:
$w=u+\xi _AV_{Ar}$. The factor $\xi _A$
describes the possible deviation of the cosmic ray drift velocity from the gas velocity.
We  use values $\xi _A=1$ and $\xi _A=-1$ upstream of the
forward and reverse shocks respectively, where Alfv\'en waves are
generated by the cosmic ray streaming instability and propagate in
the corresponding directions. The damping of these waves heats the gas
upstream of the shocks (see McKenzie \& V\"olk \citealp{mckenzie82})
and limits the total compression ratios of CR modified shocks.
%The situation is less clear in the
%downstream region and we shall consider cases $\xi _A=0$ and
%$\xi _A=-1$ downstream of the forward shock at $R_c<r<R_f$.
In the downstream region of
the forward and reverse shock at $R_b<r<R_f$ we put $\xi _A=0$ and therefore $w=u$.

We use the diffusion coefficients $D_f$ and $D_b$ at the forward and reverse shocks in the following form
\begin{equation}
 D_{f,b}=\eta _BD_B\left( 1+\frac p{P_{f,b}}\right) ,
\end{equation}
were $D_B=vpc/3qB$ is the Bohm diffusion coefficient. Since it is expected that
 the highest energy particles are
scattered by small-scale magnetic fields generated in the course
of the non-resonant streaming instability, their diffusion
coefficient  is larger  than the Bohm diffusion coefficient and is
proportional to $p^2$ (Dolginov \& Toptygin \citealp{dolginov67}).
This effect is described by the second term in parenthesis where $P_f$ and $P_b$ are the  momenta which
 separate these two different regimes of diffusion.
The parameter $\eta _B$ describes the possible deviations of
diffusion coefficient from the one achieved in the regime of Bohm
diffusion at small energies. It is expected that $\eta _B>1$ because small energy particles
can be resonantly scattered only by a
fraction of the magnetic spectrum. Throughout the paper we use
the value $\eta _B=2$.  Note that at high energies  the particle diffusion is determined by
 the small-scale magnetic fields,  and the  values of $P_{f.b}$ in Eq. (4) are adjusted to reproduce the 
 gamma and X-ray observations (see below). Therefore  the results of calculations do not strongly depend 
 on the chosen value of $\eta _B=2$.  
 
In real situations  the level of MHD turbulence drops with distance
upstream  the shock, so  the diffusion could be quite fast there. The
characteristic diffusive scale of highest energy particles is
 a small fraction ($\xi _0<<1$)  of the shock radius (see Zirakashvili \& Ptuskin
\citealp{zirakashvili08}),  and is determined by the generation and
transport of MHD turbulence in the upstream region. The value $\xi _0\sim \ln ^{-1}(D_c/D_s)$
is determined by the ratio of diffusion
coefficient $D_c$ in the circumstellar medium and diffusion coefficient $D_s<<D_c$ in the vicinity of
the shock. The MHD turbulence is amplified exponentially in time
before the shock arrival from the background
level by the cosmic ray streaming instability.
This process is not modeled  in  the present study;  we
simply multiply the CR diffusion coefficient $D$ to  the additional
factor $\exp ((r-R_f)/\xi _0R_f)$ upstream the forward shock and to  a  similar factor
$\exp ((R_b-r)/\xi _0R_b)$ upstream   the reverse shock. The characteristic range of variation of
$\xi _0$ is $0.05\div 0.1$ (Zirakashvili \& Ptuskin \citealp{zirakashvili08}).
Below  we  use the value $\xi _0=0.05$.

In  supernova remnants,  the plasma is heated to keV temperatures.
Generally, in young  SNRs the  thermal electrons are not in equilibrium with ions.  Here we assume
that the exchange between
electrons and ions  proceeds at the minimum rate, i.e. through the  Coulomb collisions (Spitzer \citealp{spitzer68}).

%%%%%%%%%%%%%%Fig.1
\begin{figure}[t]
%\figbox*{}{}{
\includegraphics[width=7.5cm]{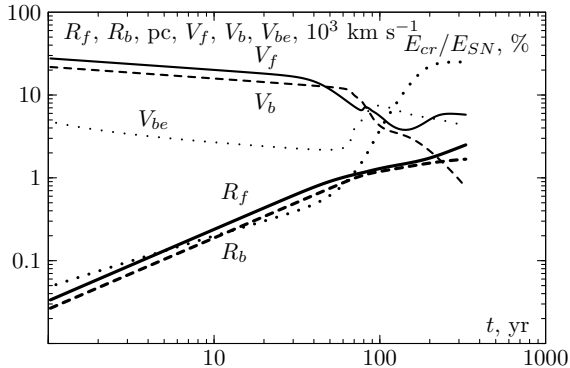}
\caption{Time dependencies of parameters characterizing the forward and reverse shocks:
the forward shock radius $R_f$ (thick solid line),
the reverse shock radius $R_b$ (thick dashed line), the forward shock speed $V_f$
 (thin solid line), the reverse shock speed $V_b$ (thin dashed line), the
reverse shock speed in the ejecta frame  $V_{be}$
 (thin dotted line). The ratio of the energy released in cosmic rays
 to the  total energy of the supernova
explosion  $E_{cr}/E_{SN}$ (dotted line) is also shown. }
\end{figure}

%%%%%%%%%%%%%%%%Fig.2
\begin{figure}[t]
%\figbox*{}{}{
\includegraphics[width=8.0cm]{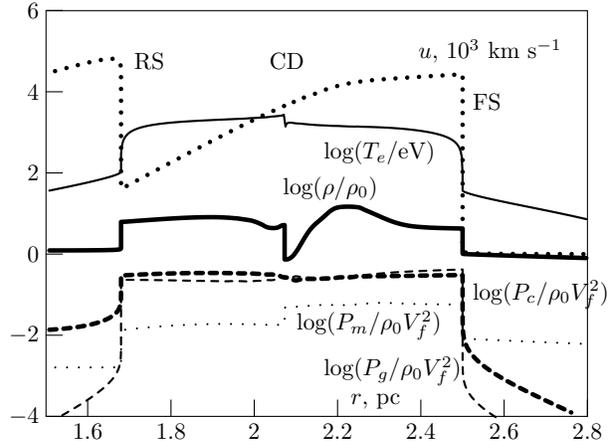}
\caption{Radial dependencies of the gas density (thick solid line), the gas
velocity (thick dotted line), the CR  pressure (thick dashed line), the gas pressure (dashed line),
the magnetic pressure (dotted line) and
the electron temperature (solid line) at the epoch
 $t=330$ yr. In the same figure the positions of the forward and reverse  shocks,
(FS and RS, respectively) as well as of the contact discontinuity (CD) are shown. }
\end{figure}

\section{Modeling of hydrodynamics and particle acceleration}

It is expected that a circumstellar medium around progenitors of
IIb supernova is strongly nonuniform. At the main
sequence (MS) phase,  the stellar wind of progenitor creates
a rarefied bubble in the surrounding medium. Later the part of this
bubble is filled with a dense gas ejected by progenitor at the Red
Super Giant (RSG) phase of the stellar evolution
(see Chevalier \citealp{chevalier05} for a review).
Typically,  the progenitors of IIb Supernova are Red-Super
Giants (RSG) which have lost almost all hydrogen envelope vie powerful
stellar wind with mass-loss rate $\dot{M}=10^{-5}-10^{-4}\
M_{\odot}$ yr$^{-1}$.  It is likely, however,  that the
progenitor  of Cas A  prior the explosion had a short
Wolf-Rayet or blue supergiant stage.  At this stage the progenitor emits  a tenuous and
fast stellar wind that creates a low  density bubble around the progenitor. The fast wind also can
compress  an inner part of the external slow RSG wind. As a result,  a narrow high-density RSG shell is formed at a distance $R_s$.

For Cas~A such a model  has been  proposed by Chevalier \& Liang \cite{chevalier89}.
It  is possible that
almost all progenitors of IIb supernova pass through this stage. Since the progenitor of  the famous
type IIP  supernova  1987A  also has been  in
the blue super giant stage,  we perhaps can extend this conclusion
to other   core-collapse supernova progenitors.

The hydrodynamical model with the circumstellar
 bubble  was also considered  for  Cas A SNR by Borkowsky et al. \cite{borkowsky96},
by Hwang \& Laming \cite{hwang09} and by Schure et al. \cite{schure08}.

Stellar evolution models for low mass IIb supernova  progenitors
(Gregory \citealp{gregory12}) do indicate  presence of the blue
supergiant stage. The blast wave of IIb SNR 1993J  has been  propagating
in the dense RSG wind one decade after explosion and later entered
into a low density region (Weiler et al \citealp{weiler07}).
The main part of the mass lost by the progenitor of
1993J supernova is still outside the SNR shell.

In any case the  forward shock of Cas A propagates into  the stellar wind.
A steady stellar wind has a  $r^{-2}$ density profile. The
mass density of  the stellar wind is given by
\begin{equation}
\rho (r)=\frac {\dot{M}}{4\pi u_wr^2}.
\end{equation}
Here $u_w$ and $\dot{M}$ are  the speed and mass-loss rate of the stellar wind respectively.
For the RSG wind speed  we  will  use the characteristic
value of $u_w=20$ km s$^{-1}$ which agrees with
the measured motion of HI absorption features observed in the direction of Cas~A
(Reynoso et al. \citealp{reynoso97}). It should be noted that most of  measurements show
that most of RSGs  have lower speeds.
However the wind speeds are higher for RSGs with high mass-loss (see Tables 1, 2 of
Mauron \& Josselin \citealp{mauron11}).
Since the forward shock speed is  much larger compared to
the wind speed,  the results depend only on the ratio $\dot{M}/u_w$.

For simplicity we assume that inside the shell at $r<R_s$ the
 gas density is given by the same Eq. (5) with the ratio
$\dot{M}/u_w$ a factor of 100
 lower in comparison with the ratio in the outlaying RSG wind.
Two regions are separated by the thin gas shell at
$r=R_s$ with
the mass equal to the mass of the swept up RSG wind. For the numerical modeling
we assume that the shell and the transition region
 have the thickness 0.1$R_s$. The actual density distribution is more complex inside
 the shell (see Schure et al. \citealp{schure08}).
However it influences only on the early evolution of SNR when the forward
shock propagates in this region. The results obtained at later epochs depend only on the shell radius
$R_s$ and the RSG wind density because of the low gas density in the bubble region.

\begin{table*}[t]
\begin{center}
\caption{Simulated models of Cas A }
\begin{small}
\begin{tabular}{|ccccc|ccccc|cccccccc|}
%{lccp{2.4cm}}
%p{0.4cm}cccp{0.4cm}{p{0.4cm}p{0.4cm}p{0.4cm}p{0.4cm}p{0.4cm}{p{0.4cm}p{0.4cm}p{0.4cm}p{0.4cm}p{0.4cm}p{0.4cm}}
%\tableline\tableline
\tableline
   model    &${M_{ej}}^a$&${k_{ej}}^b$&${R_{f}}^c$&$D^d$&${R_s}^e$&${M^f_A}^f$&${M^b_A}^g$&${n_H}^h$&${E_{SN}}^i$&${n_{ej}}^j$&${R_b}^k$&${R_c}^l$&${V_f}^m$&${V_b}^n$&${V_{be}}^o$&${B_f}^p$&${B_b} ^q$ \\
\tableline
H1          & 2.0        & 9        &  2.5      & 3.4 & 1.5     &4.5        & 8         &0.40     &1.2         &0.69     & 1.68    & 2.07    & 5.8     & 0.77     &4.2         & 1.16    & 0.57         \\
H2          & 2.0        & 9        &  2.5      & 3.4 & 0       &5          & 5         &0.36     &1.2         &0.67     & 1.82    & 1.90    & 6.0     & 4.0      & 1.4        &1.0      & 0.27        \\
L1          & 2.0        & 9        &  2.5      & 3.4 & 0       &10         & 10        &1.0      &2.0         &0.30     & 1.64    & 1.74    & 5.7     & 3.1      & 1.8        &0.78     & 0.12        \\

\tableline
\end{tabular}
\end{small}
\end{center}

$^{a}${mass of ejecta, solar masses}\\
$^{b}$ {power-law index of ejecta density distribution}\\
$^{c}${forward shock radius, pc}\\
$^{d}${distance to SNR, kpc}\\
$^{e}${RSG shell radius, pc}\\
$^{f}${parameter of magnetic amplification at the forward shock}\\
$^{g}${parameter of magnetic amplification at the reverse shock}\\
$^{h}${undisturbed hydrogen number density at the forward shock position, cm$^{-3}$}\\
$^{i}${explosion energy, 10$^{51}$ erg }\\
$^{j}${number density of unshocked ejecta, nucleons cm$^{-3}$}\\
$^{k}${reverse shock radius, pc}\\
$^{l}${radius of the contact discontinuity, pc}\\
$^{m}${forward shock speed, 10$^3$ km s$^{-1}$}\\
$^{n}${reverse shock speed in the laboratory frame, 10$^3$ km s$^{-1}$}\\
$^{o}${reverse shock speed in the ejecta frame, 10$^3$ km s$^{-1}$}\\
$^{p}${magnetic field just downstream of the forward shock, mG}\\
$^{q}${magnetic field just downstream of the reverse shock, mG}
\end{table*}

\begin{table*}
\begin{center}
\caption{Details of particle acceleration }
\begin{small}
\begin{tabular}{|cccc|ccccc|ccccc|}
%{lccp{2.4cm}}
%p{0.4cm}cccp{0.4cm}{p{0.4cm}p{0.4cm}p{0.4cm}p{0.4cm}p{0.4cm}{p{0.4cm}p{0.4cm}p{0.4cm}p{0.4cm}p{0.4cm}p{0.4cm}}
%\tableline\tableline
\tableline
 model      &${\eta _{f}}^a$&${\eta _b }^b$&${\eta ^e_{b}} ^c$ &${\eta ^e_{f} } ^d$&${P_f}^e$  &${P_b}^f  $&${p^e_f} ^g$&${p^e_b} ^h$    &${K_{en}^f}^i$  &${K_{en}^b}^j$&${E_{CR}/E_{SN}}^k$&${\sigma _f}^l$&${\sigma _b}^m$\\
\tableline
H1    & 0.01          & 0.01         &$3.5\cdot 10^{-6}$       &$2.2\cdot 10^{-6}$ &0.94       &9.4        &$p_f$       &40         &$4\cdot 10^{-4}$& 0.004        & 0.25               & 4.3          &4.9\\
H2    & 0.01          & 0.01         &$3.5\cdot 10^{-6}$       &$2.5\cdot 10^{-6}$ &$\infty $  &$\infty $  &$p_f$       &90         &$5\cdot 10^{-4}$& 0.18         & 0.18               & 4.4          &4.4\\
L1    &$10^{-7}$      & 0.01         &$3.5\cdot 10^{-6}$       &$4.0\cdot 10^{-9}$ &$\infty $  &$\infty $  &200         &200        & 0.25           & 0.73         & $3\cdot 10^{-3}$   & 4.0          &5.5\\

\tableline
\end{tabular}
\end{small}
\end{center}

$^{a}${injection rate of protons at the forward shock}\\
$^{b}${injection rate of oxygen ions at the reverse shock}\\
$^{c}${injection rate of electrons (positrons) at the reverse shock at present epoch, see Eq. (2)}\\
$^{d}${injection rate of electrons at the forward shock, for the model L1 see Eq. (1)}\\
$^{e}${momentum that separates two regimes of diffusion at the forward shock, TeV/c, see Eq. (4)}\\
$^{f}${momentum that separates two regimes of diffusion at the reverse shock, TeV/c, see Eq. (4)}\\
$^{g}${injection momentum of electrons at the forward shock, MeV/c}\\
$^{h}${injection momentum of electrons at the reverse shock, MeV/c}\\
$^{i}${electron to proton ratio at the forward shock calculated for the energy 10 GeV}\\
$^{j}${electron to nucleon ratio at the reverse shock calculated for the energy 10 GeV}\\
$^{k}${ratio of cosmic ray energy $E_{CR}$ to the energy of supernova explosion $E_{SN}$}\\
$^{l}${total compression ratio of the forward shock}\\
$^{m}${total compression ratio of the reverse shock}
\end{table*}

The RSG wind density is poorly constrained by X-ray observations of Cas A. The matter is that most of
thermal X-rays
 are produced in the shocked ejecta gas. In addition the forward shock region shows a significant
non-thermal X-ray component. That is why the flux of thermal
X-rays  and the corresponding amount of the
 stellar wind gas swept up by the forward shock are rather uncertain in this region.

On the other hand,  the ejecta mass estimate $2-4$ $M_{\odot}$ (Vink et al. \citealp{vink96}, Willingale et al.
\citealp{willingale02,willingale03}; Hwang \& Laming \citealp{hwang03}; Laming \& Hwang \citealp{laming03})
are rather robust. This range of ejecta masses is also in agreement with stellar evolution models
(e.g. Gregory \citealp{gregory12}) and the predicted light curves (e.g. Iwamoto \citealp{iwamoto97})
of IIb supernova.
We shall use the lower limit of ejecta mass $M_{ej}=2M_{\odot}$ below.

%The wind density was adjusted in order to reproduce the fluxes of X-ray and gamma-emission from Cas A.

In this study  we  use the solar composition for the stellar wind material. The corresponding
number density  of helium nuclei is $0.1n_H$ where $n_H$ is the undisturbed number density of hydrogen
 ions at the current forward shock position. So the total nucleon density is $n=1.4n_H$.

%%%%%%%%%%%%%%%%   Fig.3
\begin{figure}[t]
%\figbox*{}{}{
\includegraphics[width=8.0cm]{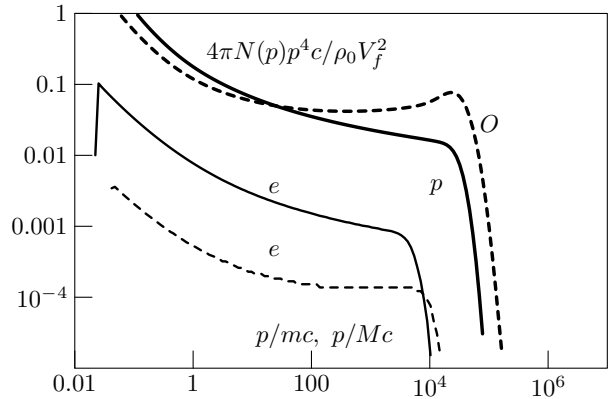}
\caption{The energy distributions of  protons at the forward shock
(thick line), of oxygen ions at reverse shock (thick dashed line), of
electrons at the forward shock (multiplied to the factor of $10^2$,
thin lines) and of electrons at the reverse shock (thin dashed line)
calculated for the model H1 at the  epoch $t=330$~yr.
Particle momenta are normalized to the proton mass $m$ and the mass of oxygen
ion $M$. The oxygen spectrum is normalized to the nucleon number
density. }
\end{figure}

%%%%%%%%%%%%%%%%%%%%%   Fig.4
\begin{figure}[t]
%\figbox*{}{}{
\includegraphics[width=8.0cm]{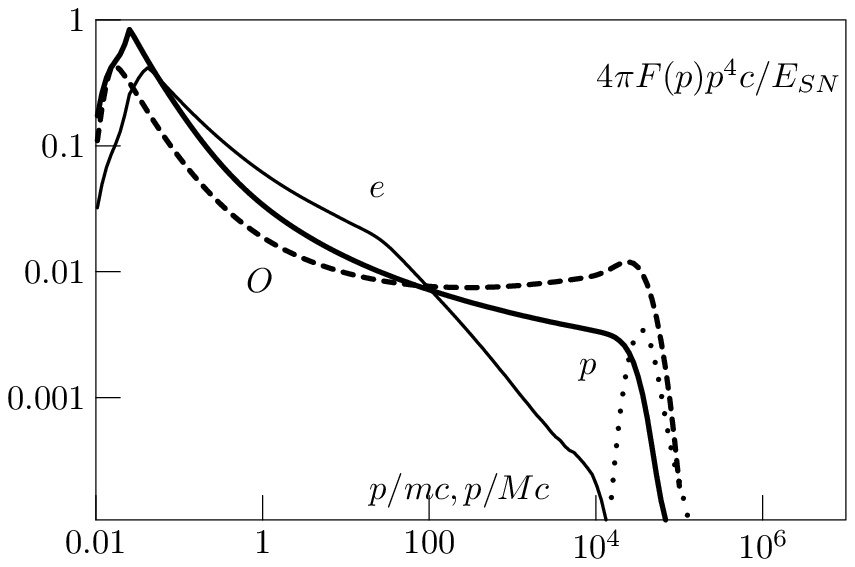}
\caption{Spatially integrated spectra of accelerated protons
(solid line), oxygen ions (dashed line) and electrons multiplied
to $10^3$ (thin solid line) at $t=330$ yr obtained in
the model H1. Spectrum of run-away protons which have left the
remnant is also  shown (dotted line). }
\end{figure}

The principal parameters used in calculations  are summarized in Table~1
and  Table~2. Both tables consist of three parts (separated by vertical lines).
The first parts contain  general parameters that are fixed for all model calculations.
The second parts contain free parameters which are  adjusted to match the key
observations. The third parts contain  the results derived from modeling.

%%%%%%%%%%%%%%%%   Fig.5
\begin{figure}
%\figbox*{}{}{
\includegraphics[width=8.0cm]{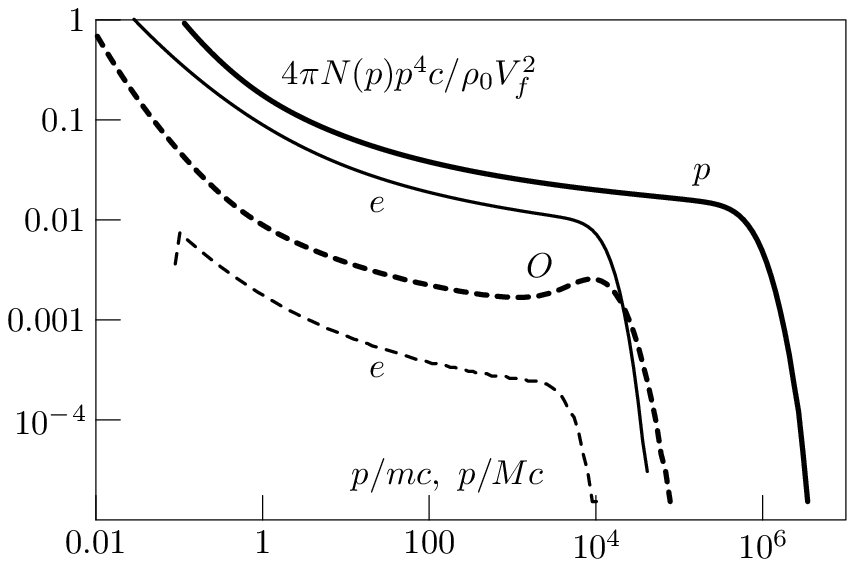}
\caption{The energy distributions of  protons at the forward shock
(thick line), of oxygen ions at the reverse shock (thick dashed line), of
electrons at the forward shock multiplied to $10^3$
(thin lines) and of electrons at the reverse shock (thin dashed line) calculated for the model H2
 at the epoch $t=330$~yr. Particle momenta are normalized to
the proton mass $m$ and the mass of oxygen
ion $M$. The oxygen spectrum is normalized to the nucleon number
density. }
\end{figure}

%%%%%%%%%%%%%%%   Fig.6
\begin{figure*}[t]
\includegraphics[width=14.0cm]{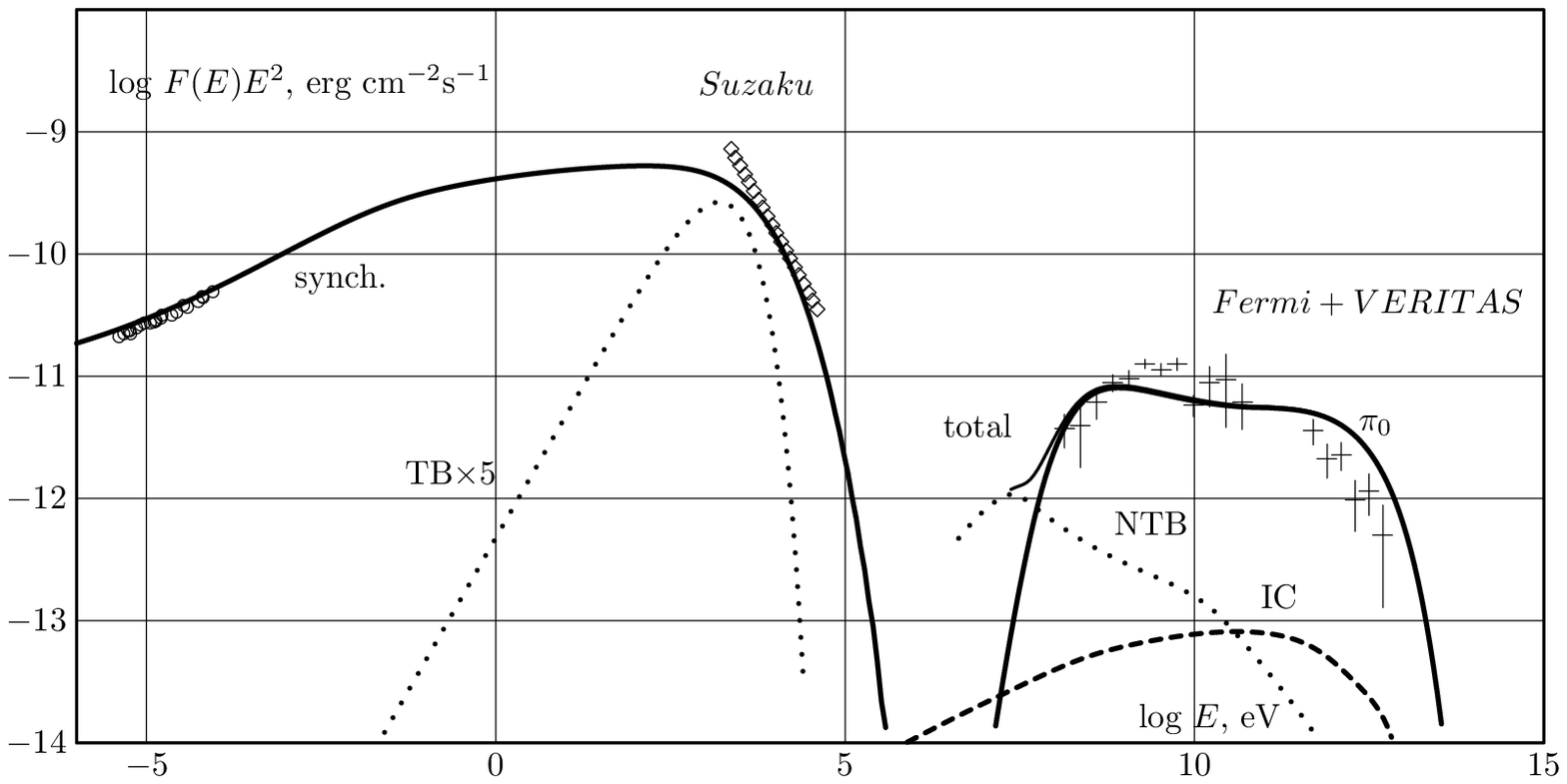}
\caption{The broad-band spectral energy distribution of
nonthermal radiation of Cas A  calculated within the hadronic model H1.
The following
radiation processes are taken into account: synchrotron radiation
of accelerated electrons (solid curve on the left), IC emission
(dashed line), gamma-ray emission from pion decay (solid line on
the right), thermal bremsstrahlung (dotted line on the left),
nonthermal bremsstrahlung (dotted line on the right).
%The input of the reverse shock is
%shown by the corresponding thin lines.
Experimental
data in  gamma-ray  (Fermi LAT, present work); VERITAS, Acciari et al. \citealp{acciari10}, data
with error-bars) and
radio-bands (Baars \citealp{baars77}, circles), as well as the  power-law approximation of Suzaku X-ray data
(Maeda et al. \citealp{maeda09}, diamonds) from the whole remnant  are also shown. }
\end{figure*}

%%%%%%%%%%%%%%%   Fig.7
\begin{figure*}[t]
\includegraphics[width=14.0cm]{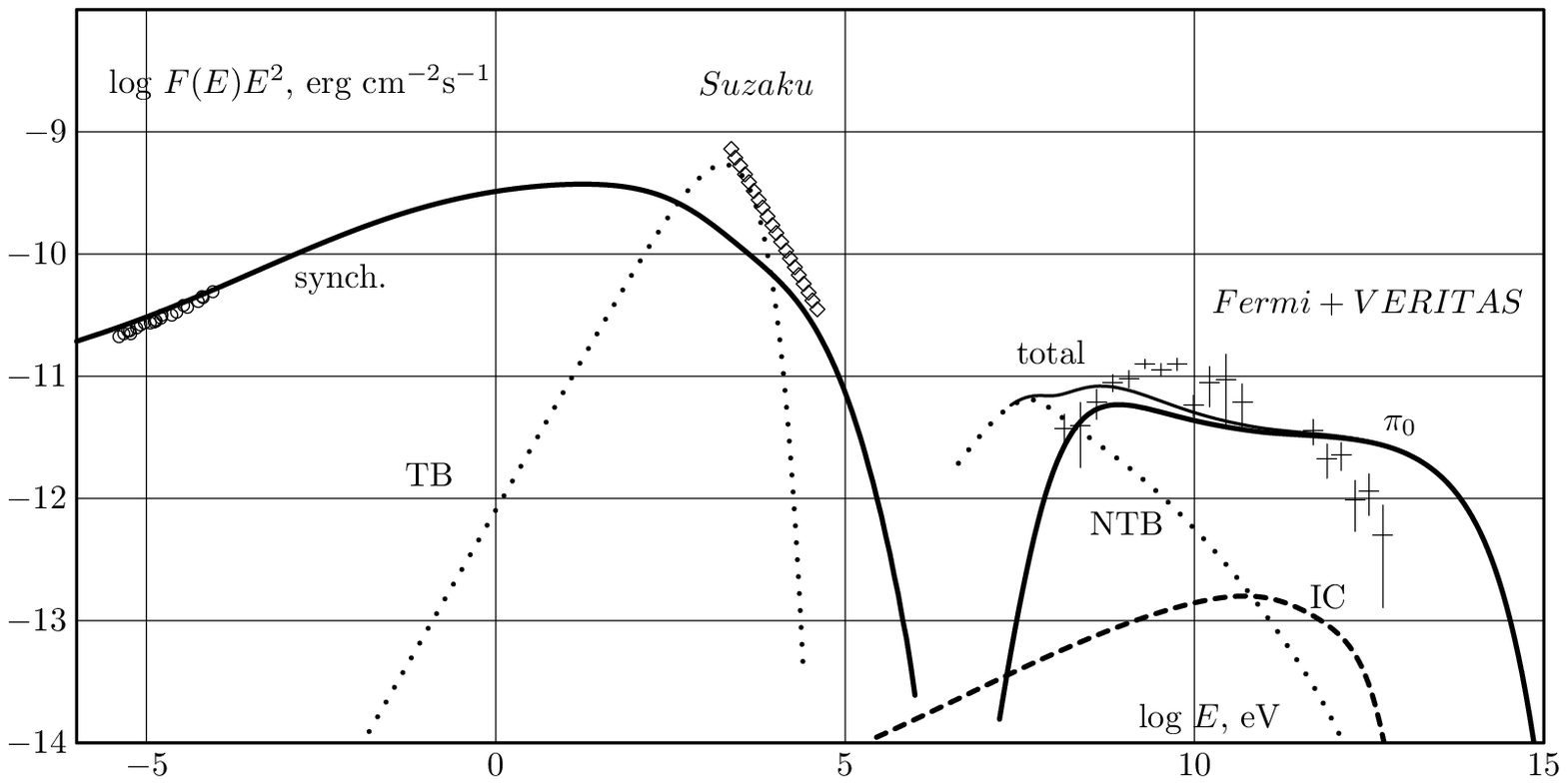}
\caption{The broad-band spectral energy distribution of
nonthermal radiation of Cas A  calculated for the hadronic model H2.
The following
radiation processes are taken into account: synchrotron radiation
of accelerated electrons (solid curve on the left), IC emission
(dashed line), gamma-ray emission from pion decay (solid line on
the right), thermal bremsstrahlung (dotted line on the left), nonthermal bremsstrahlung
(dotted line on the right).
The  detected spectral points in the gamma-ray   ({\it Fermi} LAT, this work; VERITAS, Acciari et al. \citealp{acciari10}) and
radio bands (Baars \citealp{baars77}, circles), as well as the  power-law approximation of the
Suzaku X-ray data (Maeda et al. \citealp{maeda09}, diamonds) from the whole remnant  are also shown.}
\end{figure*}

\subsection{Hadronic models}

For hadronic model H1 we use high injection efficiencies $\eta
_f=\eta _b=0.01$ at forward and reverse shocks. The stellar wind
density $n_H$ is then adjusted to reproduce the fluxes of
gamma-emission. The explosion energy is
 adjusted to reproduce the forward shock radius $R_f=2.5$ pc at the remnant age $t=330$ yr.
The value of
the amplified magnetic field is parameterized by the parameters $M^f_A=4.5$ and $M^b_A=8$ at forward and reverse
shocks respectively.
 These numbers regulates the shape of
 particle spectra and the corresponding spectra of gamma-emission.

We found that in this model with a rather low stellar wind density, $n_H=0.4$ cm$^{-3}$,  the amount of
suprathermal electrons  produced via  Compton scattering of gamma-rays from $^{56}$Co decay is not sufficient for explanation  of  the observed radio fluxes.  Therefore, in this model it is assumed that
the injection of  electrons at the forward shock is similar  to
the injection of protons.

%%%%%%%%%%%%%%%   Fig.8
\begin{figure*}[t]
\includegraphics[width=14.0cm]{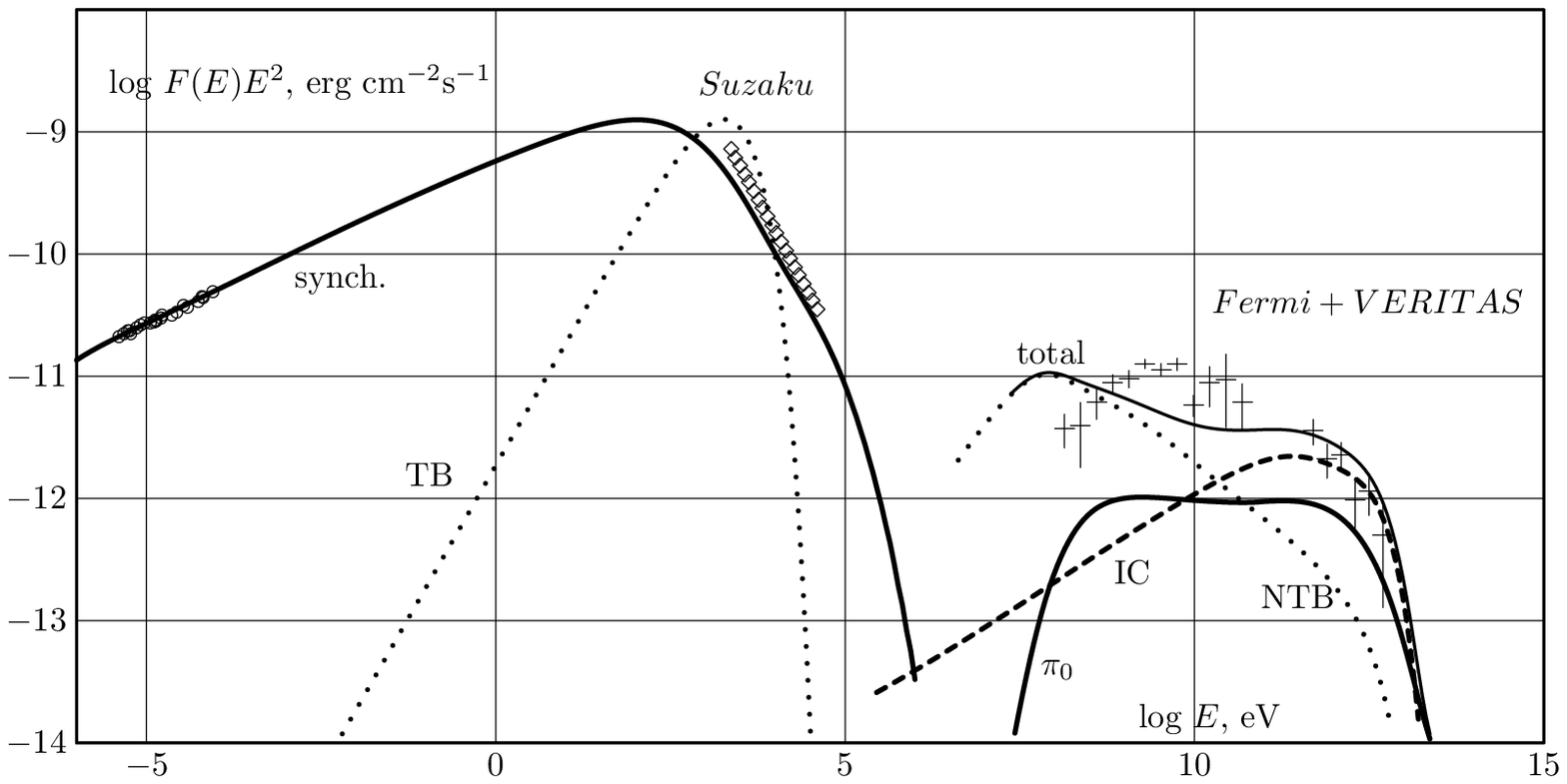}
\caption{The  broad-band spectral energy distribution  of Cas A calculated for the leptonic
model  of gamma-ray emission (model L1). The following
radiation processes are taken into account: synchrotron radiation
of accelerated electrons (solid curve on the left), IC emission
(dashed line),  $\pi^0$-decay gamma-ray emission (solid line on
the right), thermal bremsstrahlung (dotted line on the left), nonthermal bremsstrahlung
(dotted line on the right).
The  detected spectral points in the gamma-ray   ({\it Fermi} LAT, this work; VERITAS, Acciari et al. \citealp{acciari10}) and
radio bands (Baars \citealp{baars77}, circles), as well as the  power-law approximation of the
Suzaku X-ray data (Maeda et al. \citealp{maeda09}, diamonds) from the whole remnant  are also shown. }
\end{figure*}

On the  other hand the amount of electrons (positrons) from the decays of $^{44}$Ti is
sufficient for  production of   radio flux of
the reverse shock.  In calculations we adopted  the mass of $^{44}$Ti found in
observations of Cas A $M_{\rm Ti}=1.6\cdot 10^{-4}M_{\odot }$ (Renaud et al. \citealp{renaud06}).

The  electron injection efficiency at the forward shock $\eta ^e_f$ and the injection
momentum 40 MeV/c of electrons (positrons) at the reverse shock were adjusted to reproduce
the radiofluxes of the shocks.

Figures 1-7 and 9-13  illustrate the numerical results obtained in the hadronic scenario.

%The value $\xi _A=0$ was used in the downstream region.
%The combination of these parameters is chosen to explain the obtained gamma-ray fluxes
%    (see next Section).

%The initial forward shock velocity is
% $V_0=4.6\cdot 10^4$ km s$^{-1}$.

The time evolution of the shock radii $R_f$ and $R_b$,  the forward and
reverse shock speeds  $V_f=\dot{R}_f$ and $V_b=\dot{R}_b$, the reverse shock speed in the ejecta frame
$V_{be}=R_b/t-\dot{R}_b$ and  the ratio of CR energy to
the  energy of supernova explosion
$E_{cr}/E_{SN}$
%and the magnetic field strength $B_f$  downstream of the forward shock
are shown in Fig.1.

At early times after SN explosion the ratio of forward and reverse
shock radii is independent on time.  Approximately after 40~years
 after the explosion,  the forward shock starts  to interact with the dense gas of the RSG shell, the speeds of both shocks
 drop and the distance between shocks decreases. At 150 years after explosion the forward shock leaves the shell
 and enters into
 undisturbed  RSG wind. At this time the forward shock
swept up the gas mass $\sim 1.6M_{\odot }$ comparable to the ejecta mass
and the transition to the Sedov phase began.
At present the forward shock speed  is $V_f=5.8\cdot 10^3$ km s$^{-1}$ that is $15\%$
higher than the measured proper motion $\sim 5.0\cdot 10^3$ km s$^{-1}$ of the X-ray filaments
(Patnaude \& Fesen \citealp{patnaude09}).

The
reverse shock position $R_b=1.68$ pc is at the upper boundary of observed values $1.5\div 1.7$ pc. For higher ejecta
 mass $M_{ej}>2M_{\odot }$ the reverse shock is shifted to larger radii in contradiction with observations.
The reverse shock speed $V_b=0.77\cdot 10^3$ km s$^{-1}$ is in agreement with
 measured in radio speed $(1.16\pm 0.50)\cdot 10^3$ km s$^{-1}$ (Delaney \& Rudnick \citealp{delaney03})
and is lower than measured in optics speed
 $3\cdot 10^3$ km s$^{-1}$ (Morse et al. \citealp{morse04}).

Radial dependencies of  several key parameters  at the present epoch
 $t=330$ yr are shown in Fig.2.   In the same figure we show the positions  of
the contact discontinuity and the forward and reverse shocks.
 CR, gas and magnetic pressures at the forward shock are 30$\%$, 42$\%$ and 6$\%$
of the ram pressure $\rho _0V^2_f$,  respectively.
At present the forward shock have swept up 2.7$M_{\odot }$ of the stellar wind material.
The reverse shock have swept up 1.66$M_{\odot }$ of ejecta while  0.34$M_{\odot }$ of ejecta
 is not shocked yet.

The minimal electron heating by Coulomb collisions still results
in rather high electron temperatures. The shocked ejecta and the
shocked stellar wind plasma are heated up to 2.0 keV and 1.4 keV
respectively.

It should be noted that our one-dimensional calculations cannot
adequately describe the development of the Rayleigh-Taylor
instability of the contact discontinuity. In real situations
the supernova ejecta and the circumstellar gas are mixed by
turbulent motions in this region (see e.g. MHD modeling of Jun \&
Norman \citealp{jun96}).

Spectra of accelerated protons, the oxygen ions and electrons are
shown in Fig.3. At the present epoch  the maximum energy of
protons accelerated in this SNR and still confined in the shell,  is about 40 TeV; the
higher energy protons have already left the remnant.

Spatially integrated proton and electron spectra at the present
epoch  $t=330$ yr are shown in Fig.4. We also show the spectrum of
run-away particles. These particles have already left the
acceleration site  through an absorbing boundary at $r=1.2R_f$.
The sum of the proton spectra shown  is the total  cosmic ray
spectrum produced in this SNR  over the last 330 years after SN
explosion. For this SNR, the spectrum of  cosmic ray protons have
a maximum energy
 about 60 TeV that is lower than the knee in the observable cosmic ray spectrum.

To demonstrate the influence of the enhanced diffusion and RSG shell we consider
also the hadronic model H2 with the parameters similar to the parameters of the model H1 but
without RSG shell and with Bohm diffusion ($P_{f,b}=\infty $, see Tables 1,2).
Spectra of accelerated protons, oxygen ions and electrons are
shown in Fig.5.
In this model with slower diffusion the maximum energy of the protons accelerated is close to 1 PeV.

\subsection{Leptonic model}

In leptonic models of gamma-ray emission,  protons play negligible  role in the forward shock dynamics and
in the production  of electromagnetic emission. Therefore
we use  rather low injection efficiency $\eta _b=10^{-7}$ of protons at the forward shock. This very low injection
 efficiency is possible because the forward shock propagates in the stellar wind with the azimuthal magnetic field.
The magnetic field amplification here can't be produced by accelerated particles. It is probably due to vortex motions
which are created downstream of the shock in the course of an interaction of stellar wind density fluctuations with
the shock front (see Giacalone \& Jokipii \citealp{giacalone07}).

%The electron injection at the reverse shock was taken 3 times lower to suppress the input of
%IC and nonthermal bremsstrahlung emission.
%The cosmic ray pressure $P_{cr}$
%at the forward shock decreases correspondingly down to $0.14\%$ of the shock ram pressure $\rho_0 V^2_f$.
%The total cosmic ray energy also decreases up to $0.3\% $ of the explosion energy. The electron to nucleon
%ratio at the reverse shock jumps up to $K_{en}=12$. The pressure of cosmic ray positrons and electrons is close to
%$10\% $ of the ram pressure of the reverse shock.

In this model with a rather high stellar wind density $n_H=1.0$ cm$^{-3}$
the amount of suprathermal electrons generated in the stellar wind gas by the Compton
scattering of gamma-rays from $^{56}$Co decay is enough for the electron injection at the forward shock.

The electron injection efficiencies correspond to the realistic mass of $^{56}$Ni $M_{\rm Ni}=0.2M_{\odot }$ and to the
 $^{44}$Ti mass
$M_{\rm Ti}=1.6\cdot 10^{-4}M_{\odot }$ revealed by observations of Cas A
(Renaud et al. \citealp{renaud06}).
The injection momenta 200 MeV/c of electrons
from IC scattered gamma-rays
 at the forward shock and directly from $^{44}$Ti decays at the reverse shock are adjusted to reproduce radio, X-ray and gamma-ray fluxes. Since the electrons and gamma-rays from radioactive decays are produced  with small  energies,  a significant increase of their energy up to
$E \sim 200$~MeV is required. It can be realized through a stochastic mechanism of acceleration in  highly turbulent upstream regions  of  the reverse and forward shocks (Zirakashvili \& Aharonian \citealp{zirakashvili11}).   The pressure of electrons and positrons
at the reverse shock could be  comparable to  the pressure of ions.  Therefore
the reverse shock can be significantly  modified  due to
the total pressure of ions and relativistic electrons.

%%%%%%%%%%%%%%%%% Fig. 9, 10, 11
\begin{figure}
\includegraphics[width=8.0cm]{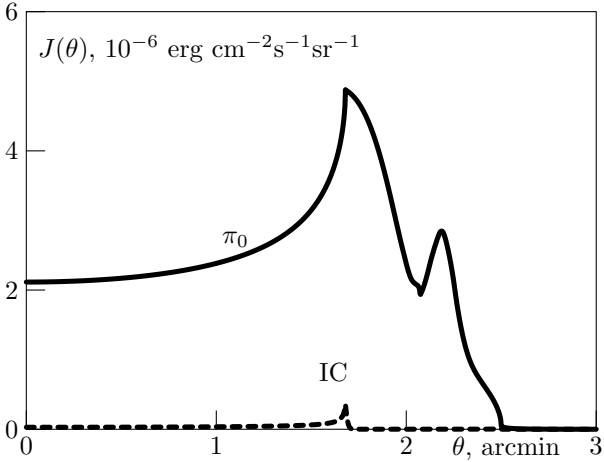}
\caption{Radial profiles of 1 TeV gamma-rays  in hadronic model H1: $\pi _0$-decay gamma-rays (solid line),
IC gamma-rays (dashed line). Gamma-emission from nonthermal bremsstrahlung is very small; its contribution is not shown.}
\end{figure}

\begin{figure}
\includegraphics[width=8.0cm]{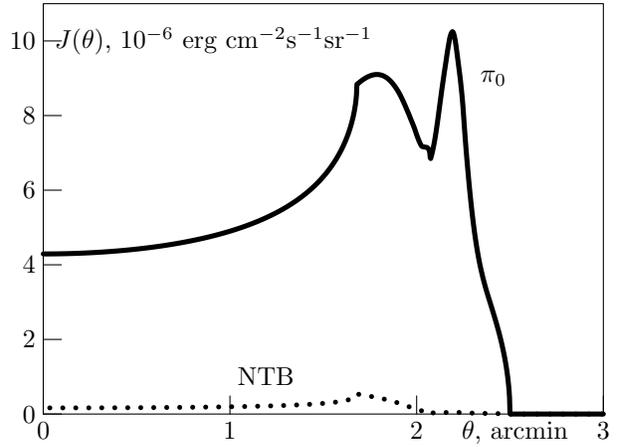}
\caption{Radial profiles of 1 GeV  gamma-rays  in hadronic model H1: $\pi _0$-decay gamma-rays (solid line),
gamma-rays from nonthermal bremsstrahlung (dotted line). The contribution of IC gamma-ray
emission is negligible.}
\end{figure}

\begin{figure}
\includegraphics[width=8.0cm]{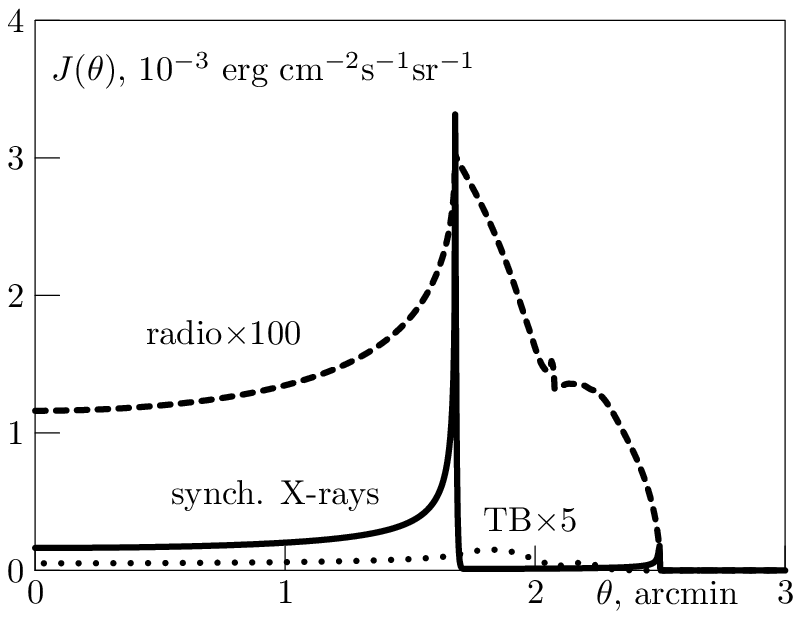}
\caption{Profiles of 5 keV X-rays  in the hadronic model H1: synchrotron X-rays (solid line), thermal
bremsstrahlung X-rays (multiplied to  the factor of 5;  dotted line).
Radio-brightness at 1.4 GHz multiplied to the factor of 100
is also shown (dashed line).}
\end{figure}

\section{Modeling of radiation}

In this section we present the results of numerical
calculations  of radiation produced in interactions of accelerated
electrons and protons. The gamma-ray spectra   from proton-proton interactions
are calculated using the formalism of Kelner et al.
\cite{kelner06}. Since $\pi^0$-decay  gamma rays from the reverse
shock are produced mainly in oxygen-oxygen collisions, we use the
following approximation. The gamma-ray flux produced by energetic
nuclei is given by  the gamma-ray flux produced by energetic
protons with a number density equal to the nucleon number density
of energetic nuclei. The target medium with a mixed  composition
is substituted by a pure hydrogen plasma with a number density
equal to the number density of nucleons in plasma with a mixed
composition. For heavy nuclei this approximation slightly
overestimate the flux of gamma-emission. This was taking into
account using the calculations of the nuclear enhancement factor
of Mori \cite{mori09}. For oxygen-oxygen collisions
 this gives the correction factor 0.6.

For IC gamma-rays we use  the standard
expressions of cross-sections (see e.g. Blumenthal \& Gould \citealp{blumenthal70}).
The main  the target photons contributing to the gamma-ray production are the microwave
background radiation and the  infra-red emission  of
Cas A itself, with the energy density 2 eV cm$^{-3}$ and temperature $T=200$ K.  The contributions of
other diffuse radiation fields can be neglected.
For calculations of synchrotron radiation we take into account that in
highly  turbulent regions close to the inverse and forward shocks,
the magnetic field has  some  probability distribution $P(B)$.
This makes the cut-off in  the spectrum of synchrotron radiation
somewhat smoother (Zirakashvili \& Aharonian \citealp{zirakashvili10a}, see also
Bykov et al. \citealp{bykov08}).

The results of  calculations of the broad-band emission for the
hadronic scenario H1 are shown  in Fig.6. The principal  model
parameters used in calculations are described in Table~1 and
Table~2. Note that at the present epoch,  25 \% of the explosion energy
($E_{SN} = 1.2 \times 10^{51}$ erg) has been  transferred to accelerated
particles (see Fig.1), most of which  are still confined in the shell of
the remnant (see Fig.4).

At low energies, the spectra of accelerated particles at  CR modified shocks become  softer
which may results in  a significant spectral steepening of  synchrotron radio emission.

The contribution of electrons directly accelerated by the
reverse shock to synchrotron radiation and IC gamma-rays is significant.  In particular,
the reverse shock produces 50$\% $ of radio and 90$\% $ of 5 keV synchrotron X-rays.
%The electron injection efficiency at the reverse shock corresponds to the  measured value
%$M_{\rm Ti}=1.6\cdot 10^{-4}M_{\odot }$.

%The energy of electron injection $E_{inj}=90$ MeV was adjusted to reproduce the observable flux of the bright
%radio-ring.

In this model gamma-rays are produced
mainly via pion decay downstream of  the forward and reverse shock (equal contributions at 1 TeV).
The modeled gamma-ray flux is slightly above the data at high energies beyond 1 TeV.
  A better agreement with data  can  be  achieved for smaller values of the parameter $\xi _0<0.05$.
% and above the Fermi LAT data below 1 GeV where gamma-rays from the nonthermal
%bremsstrahlung of electrons (positrons) from the reverse shock make  a significant contribution.

%%%%%%%%    Fig.12
\begin{figure}
%\figbox*{}{}{
\includegraphics[width=8.0cm]{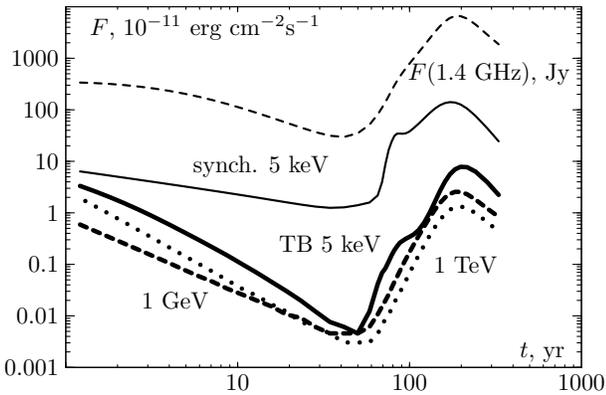}
\caption{Time dependence of fluxes of electromagnetic radiation in the hadronic model H1:
5 keV thermal bremsstrahlung (thick solid line),
5 keV synchrotron X-rays  (thin solid line),
1 GeV gamma-emission (thick dashed line), 1 TeV gamma-emission
 (dotted line); radio  at 1400 MHz (thin dashed line). }
\end{figure}

%%%%%%%%    Fig.13
\begin{figure}
%\figbox*{}{}{
\includegraphics[width=8.0cm]{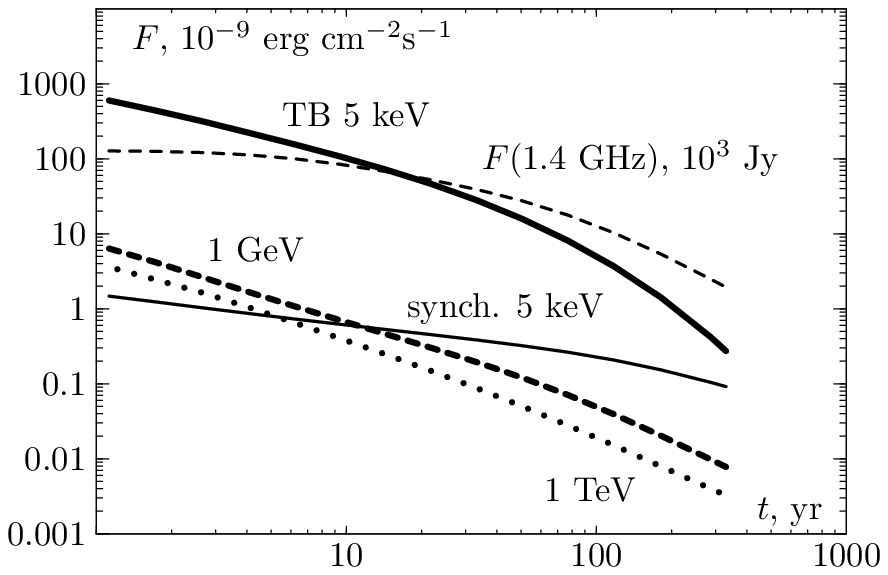}
\caption{Time dependence of fluxes of electromagnetic radiation in the hadronic model H2:
5 keV thermal bremsstrahlung  (thick solid line),
5 keV synchrotron X-rays  (thin solid line)
1 GeV gamma-emission (thick dashed line), 1 TeV gamma-emission
 (dotted line); radio  at 1400 MHz (thin dashed line). }
\end{figure}

%%%%%%%%    Fig.14
\begin{figure}
%\figbox*{}{}{
\includegraphics[width=8.0cm]{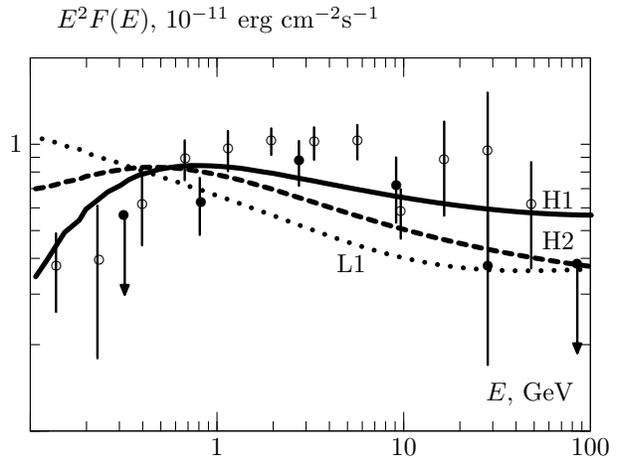}
\caption{Gamma-ray fluxes calculated within the frameworks of the models
 H1 (solid line), H2 (dashed line ) and L1 (dotted line). For comparison the fluxes
detected by  the {\it Fermi} LAT   are shown. Filled circles - spectral points reported by the Fermi collaboration (Abdo et al. \citealp{abdo10}), open circles - the results of our analysis
of the {\it Fermi} LAT data.}
\end{figure}

In Fig.6 we show also the energy flux of the  thermal bremsstrahlung. It is known that the
thermal X-rays of Cas A are produced mainly in the dense knots of supernova ejecta. To take this effect into
 account and to explain soft X-ray data we multiply the X-ray thermal bremsstrahlung flux
on the factor of $\xi _X=5$.
This means that the volume factor of the X-ray emitting plasma is below 0.2.
The flux of non-thermal X-rays is about 70$\% $ of the total X-ray flux at 5 keV.

%It has a
%maximum at 2 keV and is higher than the flux of nonthermal X-rays for photon energies below 10 keV.

The results of  calculations of the broad-band emission for hadronic model H2
are shown  in Fig.7.

In this model the ejecta gas is compressed into the thin shell downstream of the reverse shock. The thermal
 bremsstrahlung from this shell makes the main X-ray contribution at energies below 10 keV. At higher energies
 the synchrotron X-rays from the forward shock dominate in X-ray spectrum. The reverse shock produces only 12$\% $
 of non-thermal X-rays. The calculated gamma-ray emission is produced mainly at the forward shock and
is above the data points at high energies above 1 TeV.
The same is true at low energies below 1 GeV where the non-thermal bremsstrahlung of electrons accelerated at
the reverse shock gives  a non-negligible contribution. The  shocks produce equal amount of radio-emission.

The results of  calculations of the broad-band emission for leptonic scenario
are shown  in Fig.8.

The reverse shock produces 40$\% $ of radio and 50$\% $ of 5 keV X-rays.
The radio-spectrum of the non-modified
forward shock is rather steep. This is because the electrons were
mainly injected and accelerated at early stages of the remnant
expansion (see Eq.(1)) when the synchrotron losses result in the
steepening of electron energy distribution.

In this model gamma-rays are produced mainly by non-thermal
bremsstrahlung and IC scattering at the reverse shock. The modeled
gamma-ray flux is in good agreement with observations at high
energies
 but is significantly higher than Fermi LAT data below 1 GeV.

\section{Radial profiles}

The radial profiles of brightness distributions of  X-ray, gamma-ray  and radio- emissions
calculated for the hadronic model H1 are shown
in Fig.9-11. The projection effect is  taken into account.

All three components of radiation in the radio, X-ray and
gamma-ray bands  peak at the contact discontinuity and at the reverse
shock (approximately $1.7-2.0'$ of the angular radius).

The narrow peak of gamma-emission  at angular radii $\theta >2'$ (see Figures 9 and 10)
 is produced by
the forward shock protons in the gas of the RSG shell swept up by the forward shock.

The radio and X-ray profiles  in Fig.11 are in agreement with available
observations (Anderson et al. \citealp{anderson91}, Gotthelf et
al. \citealp{gotthelf01}, Helder \& Vink \citealp{helder08}).

The sharpest  features in the radial profiles appear in  X-rays (see Fig.11). Such sharp features
imply   filamentary structures  at forward and reverse shocks.
Note that the projection effect works differently for
reverse shock in comparison with the forward shock (see
Zirakashvili \& Aharonian \citealp{zirakashvili10a}). As a result,
the filaments look  similar in spite of a relatively weak
magnetic field at the reverse shock.

 The calculated width  of  filaments  shown in Fig.11 is close to $1''$ which is  slightly 
less  than the thickness measured     by {\it Chandra} (Bamba et al. \citealp{bamba04}). 
 It has been argued by  V\"olk et al. \cite{voelk05} that the observed thickness 
 corresponds to the downstream magnetic field of about 500 $\mu $G.  The larger fields  demanded by our 
 model  results in narrower filaments. The observed  broader filaments can be explained by   the  limited angular 
 resolution and/or  by inhomogeneous circumstellar medium.  

%In addition a filament of thermal X-rays is also presents in
%Fig.9. It is because the gas of ejecta heated by the reverse shock
%have very high density near the contact discontinuity. Although
%the mass density of ejecta corresponds to $n_H=2$ cm$^{-3}$ just
%downstream the reverse shock, the mass density of ejecta increases
%toward the contact discontinuity up to $n_H\sim 100$ cm$^{-3}$
%(see Fig.2).

In realistic  scenarios,  the corrugated shock surfaces and the
development of Rayleigh-Taylor instability near the contact
discontinuity will destroy a picture of  ideal circle filaments.
Nevertheless,  non-thermal X-ray filaments must be
observed in the reverse shock region in Cas A.

The dominance of ejecta in thermal X-rays is not surprising since
Cas A is at transition from the ejecta dominated stage to
Sedov stage now. It is well established that thermal X-rays from
very young extragalactic supernova are produced by a shocked
ejecta gas (see e.g. Chevalier \& Fransson \citealp{chevalier06}).

\section{Secular evolution of fluxes}

The calculated temporal dependencies of  radiation fluxes in
radio, X-ray, and gamma-ray  bands in the hadronic models H1 and
H2 is shown in Fig.12 and 13. Results for epochs $t<30$ yr shown
in Fig. 12 can be used also for Ib/c SNRs with tenuous winds of
Wolf-Rayet stars. Results for H2 model (Fig. 13) are more
appropriate for IIb SNRs with dense RSG stellar wind.
The rate of the secular decrease of electromagnetic radiation calculated
for these two hadronic models, as well as for the leptonic model,  are  summarized in Table 3.

The  rates of decrease of  GeV and TeV gamma-ray flux predicted by hadronic models are significantly
larger than the one expected in the leptonic model. It should be noted that
since the radiation field of Cas A is dominated by its own infra-red radiation,
the secular decrease of inverse Compton gamma-rays is affected by the
decrease of the  density of the target field. For simplicity, the decrease  rate of  IC
 gamma-rays  0.12$\% $ yr$^{-1}$   in  Table 3  has been calculated
assuming a time-independent radiation field. Therefore the secular decrease of IC gamma-rays
should be somewhat  faster given the secular evolution of the infra-red radiation, but still it remains
significantly slower than the decrease of gamma-ray flux in hadronic models.

The current secular decrease of radio-intensity 0.85$\% $ yr$^{-1}$ is close to
measurements of Baars et al. \cite{baars77} (0.93$\pm $0.05$\% $ yr$^{-1}$ at 1.4 GHz) and higher than measurements
 of  O'Sullivan \& Green \cite{osullivan99} (0.6$\% $ yr$^{-1}$ at 15 GHz).
At the same time, the calculated  rates of decrease of 5 keV  X-ray fluxes  predicted by hadronic and
leptonic models  close to 1\% per year  is slower than the measured value
$(1.5\pm 0.17)\% $ yr$^{-1}$ (Patnaude et al. \citealp{patnaude11}).

The fast decrease of radio and X-ray emissions
in the models H1 and L1 is explained by the significant contribution of the
reverse shock to the radio and X-ray production.
Presently, it propagates through  the flat part  of the ejecta spatial density
distribution. The ejecta density $\rho _{ej}$ drops
proportional to $t^{-3}$, i.e,  0.9\%   per year.  The  fast drop is the main
 factor which determines the decrease of thermal and nonthermal X-rays.

Since the synchrotron X-rays
are produced in the loss-dominated regime the X-ray flux $F(E)$ of the reverse shock can be written as
\begin{equation}
F(E)\propto E^{-2}\rho _{ej}V^3_{be}R_b^2f(E/E_0)
\end{equation}
where $E_0$ is the maximum energy of synchrotron X-rays and the function $f(x)$ describes
the shape of the spectral
 synchrotron cut-off  (see also Patnaude et al. \citealp{patnaude11}). 
 It is  also assumed that the the electron energy density is proportional
 to the ram pressure of the shock $\rho _{ej}V^2_{be}$.
 For the energy $E_0$  which is proportional to the square of the shock
speed $V_{be}$ , we can estimate the secular decrease of X-ray flux as
\begin{equation}
\frac {\dot{F}}{F}=\frac {\dot{\rho }_{ej}}{\rho _{ej}}+
(2\Gamma _X -1)\frac {\dot{V}_{be}}{V_{be}}+2\frac {\dot{R}_b}{R_b}
\end{equation}
Here $\Gamma _X$ is the spectral slope $\Gamma _X=-d\ln F /d\ln
E$. Using dependencies $V_{be}\sim t^{-0.6}$, $R_{b}\sim t^{0.2}$
(see Fig.1) and the observational value $\Gamma _X=3$ we obtain
$F\sim t^{-5.6}$  that is the secular decrease $1.7\% $ yr$^{-1}$.
The total
 secular decrease of X-rays is lower because the thermal X-rays and the forward shock
 give an additional contribution.
 The X-ray flux of the forward shock  decreases slower because in the stellar wind
 the density at the forward shock $\rho _0R_f^2=\mathrm{const} $. That is why
 the decrease of synchrotron X-rays at the forward shock depends only on the decrease of the forward shock
 speed. Then the equation similar to Eq. (7) gives the synchrotron flux decrease  proportional to $V_f^5$. For the
 measured dependence $V_f\sim t^{-0.2}$,  we obtain the secular decrease  at the forward shock of order of 
 $0.3\% $ yr$^{-1}$.  

This is demonstrated
in Fig. 13  for the H2 model in which the
non-thermal X-rays are produced at the forward shock.

In this regard the fast decrease ($>1\% $ yr$^{-1}$) of hard X-ray flux reported  by Patnaude et al. \cite{patnaude11}
is in favor of the interaction with the RSG shell in past (model H1). During this interaction the reverse shock speed
 $V_{be}$  increases significantly, but presently this speed decreases (see Fig.1). This results in the secular decrease
 of the non-thermal X-ray flux from the reverse shock that is faster than secular decrease of the ejecta density
 $0.9\% $ yr$^{-1}$.
 The situation is different in the models without
 RSG shell (H2, L1). In these models, the speed $V_{be}$  presently is almost constant.
 Correspondingly, the secular
 decrease of the nonthermal X-ray flux cannot  be faster  than $0.9\% $ yr$^{-1}$.

 We conclude that the model H1 is in better  agreement with the measured secular decrease, although for
X-rays even the model H1 predicts slower decrease than the observations show.

%Since non-thermal X-rays
% are produced in the energy loss dominated regime the X-ray flux of the reverse shock is proportional to
% $\rho _{ej}V_{be}^3R_b^2$. The reverse shock speed  $V_{be}$ drops approximately as $t^{-0.7}$.

We may directly compare the results of our calculations with
radio-observations of well-studied extragalactic IIb supernova
1993J (Weiler et al. \citealp{weiler07}). Its radio flux at $\lambda
=20$ cm was $\sim $ 30 mJy at day 3000. This corresponds to $\sim
$ 30 kJy recalculated for Cas A distance which is  a factor
of 3 smaller than  the value corresponding to the radio-curve in hadronic model H2 (see Fig.13).
The reason could be that in Cas A the  electron injection is more efficient in
Cas A than in SNR 1993J or it was lower in past.

The model H2 predicts gamma-ray 
flux  one year after the explosion  at the level of $10^{-8}$ erg cm$^{-2}$s$^{-1}$. 
This corresponds to the flux $\sim 10^{-14}$ erg cm$^{-2}$s$^{-1}$ recalculated for SNR 1993J distance.
This estimate agrees with  the  results of Tatischeff \cite{tatischeff09}, but is below by two orders of magnitude 
 the flux  predicted by Kirk et al. \cite{kirk95}. The reason of such a large discrepancy  
is  the higher wind density assumed by Kirk et al. \cite{kirk95}.

\begin{table}
\begin{center}
\caption{Secular decrease of electromagnetic radiation, $\% $ yr$^{-1}$}
\begin{small}
\begin{tabular}{|ccccc|}
%{lccp{2.4cm}}
%p{0.4cm}cccp{0.4cm}{p{0.4cm}p{0.4cm}p{0.4cm}p{0.4cm}p{0.4cm}{p{0.4cm}p{0.4cm}p{0.4cm}p{0.4cm}p{0.4cm}p{0.4cm}}
%\tableline\tableline
\tableline
 model      &1.4GHz&5 keV&1 GeV& 1 TeV\\
\tableline
H1    & 0.85 &1.14 & 0.70& 0.82\\
H2    & 0.53 &0.77 & 0.35& 0.47\\
L1    & 0.73 & 0.91&0.14 & 0.12\\

\tableline
\end{tabular}
\end{small}
\end{center}
\end{table}

\section{Discussion}

\subsection{Circumstellar medium around Cas A}
According to
the recent proper motion X-ray measurements (Patnaude \& Fesen
\citealp{patnaude09}), the expansion parameter of Cas A is
$m_{exp}=V_ft/R_f=0.65$   This is close to the expansion parameter
$m_{exp}=2/3$ at the Sedov phase for a supernova shock moving
in the medium with  the $r^{-2}$ density profile. At first glance  this can be interpreted
that Cas A is  already in the Sedov  phase. However this
contradicts to the current radius of the reverse shock  $R_b\sim
0.6-0.7R_f$.  Detailed hydrodynamical  simulations  confirm  this
qualitative argument. In order to reproduce the current positions of the forward
and reverse shocks in Cas A,  one has to assume for the forward shock velocity
$V_f=5600$ km s$^{-1}$ (e.g. Chevalier \& Oshi
\citealp{chevalier03}; Patnaude \& Fesen \citealp{patnaude09}).
This is close to the value $V_f=5700$ km s$^{-1}$ found in our leptonic model
calculations and is 15$\%$ higher than the proper motion
$V_f=4900$ km s$^{-1}$ of X-ray filaments.

Patnaude and Fesen \cite{patnaude09} proposed an explanation of this discrepancy assuming
that the forward shock is modified by the cosmic ray pressure. Then
the highest energy particles leave the remnant and take
away some part of energy. This results in a smaller  shock velocity.
However, this scenario is possible only if
the number density of the circumstellar medium and the ejecta mass are
several times lower compared to  the parameters
$n_H\sim 1$ cm $^{-3}$ and $M_{ej}\sim 2-4M_{\odot}$ derived from X-ray observations.
For number density  $n_H\sim 1$ cm $^{-3}$  assumed by Patnaude \& Fesen,
the gamma-ray flux form the
pion decay  would be a factor of 10 higher in comparison to
the one measured by {\it Fermi} LAT at GeV energies and a factor of 30
higher than the flux of TeV gamma-rays measured by HEGRA, MAGIC and VERITAS.

The lower number densities are indeed used in our models H1 and  H2
with an efficient cosmic ray acceleration. However strong magnetic
fields assumed results in the steep spectra of accelerated
particles. That is why the energy flux of escaped particles is low and
does not produce any effect on the forward shock velocity.

An alternative  way to resolve the problem would be  a deviation from the $r^{-2}$
density profile. The Cas A progenitor might emit
fast and tenuous wind prior the explosion and produce  a bubble of
 rarefied gas and  a shell of compressed RSG wind in its vicinity. In our hadronic  scenario,
 the collision of the forward shock with this shell results in a significant transfer of the explosion energy to the
reverse shock. Correspondingly, the energy related to the forward shock will be  reduced  leading to
a lower speed of the forward shock.
Our calculations show that in this way one can bring
the forward shock speed  in accordance with measurements for
the shell radius $R_s=1.8$ pc.  On the other hand, for this shell radius  the reverse shock radius
appears  too large compared to observations.  In our modeling we adopt an ad-hoc
(smaller) shell radius, $R_s=1.5$ pc. In this regard we should note that
it is possible to reproduce both the measured shock speed and the reverse shock radius
for higher stellar wind densities. A similar conclusion has been made
by van Veelen et al. \cite{veelen09}
who investigated the impact of a possible Wolf-Rayet phase
for the Cas A progenitor on the SNR evolution.

Severe constrains on the radius of the RSG shell have been  found by Schure et al. \cite{schure08}. They
 showed that the jet observed in Cas A can not pass through the RSG shell if its radius 
$R_s\ge 0.3-0.4$ pc.
However the stellar wind density in our model H1 is a factor of 3 lower than the stellar wind density
 assumed by Schure et al. \cite{schure08}. Therefore the jet can survive even for larger size of the bubble.

The earlier proper motion measurements by  Delaney and Rudnick \cite{delaney03}  
is in a good agreement with the forward shock speed derived from our model calculations. However one should note that 
the results of  Delaney and Rudnick \cite{delaney03} 
imply   $15-20\% $ larger  speeds of  X-ray filaments than the ones  reported by Patnaude and Fesen \cite{patnaude09} 
 (see Table 1 of Patnaude and Fesen \citealp{patnaude09}). 
 The reason of the discrepancy  remains unclear.

Finally, the low measured velocity of X-ray filaments may be related with density disturbances in the
 circumstellar medium
when we observe the brightest X-ray filaments which are formed in
the regions  with a higher stellar wind density and a
correspondingly lower  local shock speed. This scenario can be
checked easily. Patnaude \& Fesen \cite{patnaude09}  reported the
increase of the brightness of the outer X-ray filament in the
north-east periphery of Cas A. The measured value
 of its speed is $\sim $ 5000 km s$^{-1}$. Probably the forward shock recently has entered a
 denser region in this position and initiated an increase of the  X-ray brightness.
 After the shock exits  this high density region
 and  enters a lower density region,  in future we will observe a decrease of the X-ray brightness
 and increase of the filament speed.
 Taking into account that the speeds of different
 forward shock filaments show a significant scatter $1600$ km s$^{-1}$
 (see the earlier measurements of Delaney \& Rudnick \citealp{delaney03}) this scenario seems rather plausible.

We should note that our estimate of the stellar wind mass $2.7M_{\odot }$ swept
up by the forward shock in the hadronic model H1 is several times smaller than the
estimate $9M_{\odot }$ of Vink et al. \cite{vink96}. However, as indicated in Sect. 4,
 the fraction of thermal X-rays produced in the shocked circumstellar medium of Cas A is rather uncertain.
 For example  Favata et al.
\cite{favata97}, after subtraction of the  non-thermal fraction of X-ray emission,
derived $5 M_{\odot}$ of swept up  material in Cas A  which is closer to our estimate.

%Our estimate of the swept up stellar wind mass corresponds to the mass-loss rate
%$\dot{M}=1.7\cdot 10^{-5}M_{\odot }$  yr$^{-1}$ for the wind speed $u_w=20$ km s$^{-1}$. The
% ejecta mass range $M_{ej}\sim 2-4M_{\odot }$ corresponds to the expected initial mass of the progenitors
% $12-23M_{\odot }$. The expected duration of the RSG stage is close to million years for such stars.
%Then
% the total mass-loss during this stage is 17$M_{\odot }$. This number is comparable or higher than the
% expected progenitor mass.
%  The ejecta mass $M_{ej}=2M_{\odot }$ and the explosion energy $E_{SN}=10^{51}$ erg
%in our model H1 are close to the corresponding parameters in the model 13H1 of the Iwamoto et al.
% \cite{iwamoto97} obtained for the explanation of  the light curve of
%1993J supernova. The initial mass of the progenitor is 13 $M_{\odot }$ in this model.
% Only $9M_{\odot }$ of this mass were lost via stellar wind before explosion while the rest went into the neutron
% star and the supernova ejecta.
% This means that either the actual mass-loss was lower or the duration of RSG stage was shorter.
% We conclude that the mass-loss rate inferred from the rather low wind density in our hadronic models H1, H2
% does not contradict to the stellar evolution models of the Cas A progenitor.
% On the other hand higher stellar wind densities and correspondingly higher mass-loss rates
%are not in agreement with the stellar evolution models.

Lee et al. \cite{lee13} recently reported the observation of thermal X-rays produced in the shocked circumstellar
 gas beyond the
 non-thermal X-ray filaments of Cas A. According to their explanation this X-ray emission is produced at the outlaying
 parts of the forward shock where CR acceleration is not effective. They also estimated the stellar wind density and
  derived $6M_{\odot }$ of the wind
 material swept up by the forward shock. In our opinion,  it is more natural to relate the thermal
 X-ray emission to  the strong gas heating in the upstream regions of the forward shock. The Alfv\'en
 velocity is of the order of  $1000$ km s$^{-1}$ in this region (see Section 8.3 below). The corresponding
 magnetic amplification is produced by the plasma motions with similar velocities of the gas. These motions
 unavoidably  produce multiple small-scale  shocks and the gas heating in the upstream region. It is known that
 the development of the non-resonant streaming instability results in the magnetic  amplification and
 in the very complex density structure when the main part of the volume is filled with almost empty gas cavities.
 The cavities are separated by the walls with a denser gas (see e.g. Bell \citealp{bell04}). Therefore 
 mean plasma  density derived from X-ray observations can be significantly overestimated.

As for the thermal X-rays produced by ejecta,  the flux of thermal X-rays is
underproduced in our model H1 (see Section 5). The enhancement factor $\xi _X=5$ assumed
to reproduce the soft X-ray data
may be used to estimate the properties of X-ray emitting knots. The  number density of nucleons
$n$  in  the X-ray emitting plasma  is
 estimated as $n\sim n_bV_k^2/V^2_{be}$,  where $V_k$ is the shock speed in the knot, $V_{be}$ is the reverse shock
 speed in the frame of ejecta, and $n_b$ is the mean density of the shocked ejecta.
 In the model H1 we have  $V_{be}=4.2\cdot 10^3$ km s$^{-1}$,  while the shock speeds in the X-ray
 emitting knots are of the order of 800 km s$^{-1}$. This gives  $n\sim 30n_b$. Then the
 volume factor of the X-ray emitting plasma must be $\sim 1/180$ in order  to reproduce the enhancement factor $\xi _X=5$.
The
 X-ray emitting  plasma contains $\sim 1/6$ of the shocked ejecta mass. Using $n_b\sim 4$ cm$^{-3}$
 in the model H1,  we estimate
 the nucleon number density of the X-ray emitting plasma $n\sim 120$ cm$^{-3}$. The corresponding electron number
 density $n_e\sim n/2=60$ cm$^{-3}$ appears to be  in agreement with the results of Lazendic et al. \cite{lazendic06}.
 The
 measured mean
 ionization age $n_e\tau \sim 2\cdot 10^{11}$ cm$^{-3}$ s (Lazendic et al. \citealp{lazendic06},
 Hwang \& Laming \citealp{hwang12}) then corresponds to the age of the X-ray emitting plasma $\tau \sim 100$ years.

In the model H1, the ram pressure of the stellar wind  is $\rho _0u_w^2\sim 3.7\cdot 10^{-12}$ erg cm$^{-3}$.
This value is comparable with the pressure of the interstellar medium. It is expected that the RSG wind is embedded
 in the rarefied bubble created by the fast wind of Cas A progenitor at the main sequence stage.
 The gas in the bubble is
 probably in the pressure equilibrium with the interstellar medium.
Then the RSG wind is terminated somewhere not
 far from the current forward shock position.  This occurs at distances $3.5-5$ pc
 for a range of interstellar pressures $1-2\cdot 10^{-12}$ erg cm$^{-3}$.

The stellar wind gas is compressed and forms a dense outer
 shell (see modeling of P\'erez-Rend\'on et al. \citealp{perezrendon09}).
The forward shock of Cas A will encounter this shell in future.
The expected mass of RSG wind material beyond the current
 forward shock position including the outer RSG shell  cannot be less than $5M_{\odot }$.
Another high density shell with
 mass $\sim 10^4 M_{\odot}$
 forms the walls of the rarefied  bubble at distances several tens of parsecs
(P\'erez-Rend\'on et al. \citealp{perezrendon09}). Some gas shells are indeed observed in infrared in the
 vicinity of Cas A (Barlow et al. \citealp{barlow10}).

Both shells are considered as perspective targets for
 gamma-ray production by the highest energy protons escaping SNR Cas A.
 These protons had no time to travel far
 from the SNR because even a rectangular propagation at 300 years corresponds to 100 pc distance.
Probably they are captured by tangled magnetic fields in the bubble walls.

\subsection{Comparison of hadronic and leptonic models}

The gamma-emission calculated in the hadronic model H1  is in reasonably good agreement
with measurements  (see Figure 6). The steepening of the gamma-ray spectrum at TeV energies
is achieved by assuming  faster diffusion of highest energy particles in accordance with
 Eq.(4). This assumption  affects only the  diffusion of highest energy protons and nuclei, but
 does not  have a strong impact  on electrons;  below the maximum energy of electrons around
10-20 TeV determined by synchrotron losses  (see Figure 3) the diffusion is still close to the
Bohm limit. The adjusted values of momenta $P_f= 0.9$, $P_b= 9$ TeV/c
 which
 separate two regimes of diffusion can be neither increased nor decreased. The increase of
$P_{f,b}$ will lead to  contradiction with TeV gamma-ray data,  while the decrease of $P_{f,b}$ will contradict
to
 hard X-ray data. Thus we can conclude that the TeV gamma-ray and hard X-ray emission
components are produced by particles scattered by small-scale magnetic inhomogeneities
(see Eq. (4)).
 %For example on can try  to use the higher value of $\eta _B=20$ with Bohm diffusion ($p_0=\infty $) to explain the spectrum of gamma-emission at high energies. However this will decrease the maximum energy of non-thermal X-rays by a factor of 10  and will result in contradiction with X-ray data.

%Without the faster diffusion (model H2) the gamma-emission is significantly overproduced at energies above 1 TeV (see Fig. 7).

Diffusion coefficient corresponding to  the scattering by
small-scale magnetic inhomogeneities is estimated as
 $D=vkr^2_g/3$. Here $r_g=pc/qB$ is the gyroradius of a charged particle calculated in the small-scale random magnetic field
 with the square root of the mean square $B$, $k$ is the inverse scale of the random field. If the magnetic field is
 amplified in the course of the non-resonant streaming instability,
then $k=\gamma /v_A$ (Bell \citealp{bell04}).
Here $\gamma $ is the increment of the
instability and $v_A$ is the Alfv\'en velocity calculated in the non-amplified magnetic field. Comparing the diffusion
 coefficient with  Eq.(4),  one finds  $P_f=\eta _BqBv_A/c\gamma $. We may estimate the increment of
the instability as $\gamma \sim 10\ t^{-1}$ where $t$ is the age of  the SNR.
The value of the amplified field at the forward shock is
$B=V_f\sqrt{4\pi \rho _0}/M_A^f$ (cf. Eq. (3)).  Finally we obtain

\[
cP_f=\frac {\eta _BqBv_A}{\gamma }\sim \frac {\eta _BqV_ftB_0}{10M_f^A}=
\]
\begin{equation}
\mathrm{95\ TeV\ } \eta _B\frac {t_{\mathrm{kyr}}B_{0,\mu \mathrm{G}}}{M^f_A}\left( \frac {V_f}{10^3 \mathrm{km\ s}^{-1}}\right)
\end{equation}
Here $B_0$ is the magnetic field in the stellar wind. For the forward shock
$M_A^f=4.5$, $V_f=5.8\cdot 10^3$ km s$^{-1}$ and
 the adjusted value of $P_f$ corresponds to the magnetic field strength in the RSG
stellar wind $B_0=0.012 \mu$G.
It corresponds to the magnetosonic
Mach  number of the wind $M_w=u_w/v_A\sim 570$, while the upper limit $\sim 7\mu$G
corresponds to the condition $M_w=1$.
 In reality,  the
 stellar wind with an  azimuthal magnetic field and $M_w<5$ is collimated close to the parent star.
 Since the forward shock in  Cas A is rather circular,  we conclude that the wind collimation does not occur
 and the magnetosonic Mach number $M_w>5$. Then our estimate of the wind magnetic field seems reasonable.
 The scale of the amplified magnetic field is only $k^{-1}=v_A/\gamma \sim 10^{13}$cm, i.e. $10^{-6}$
 fraction of the remnant radius. It is comparable with the
 gyroradius of TeV protons calculated in the amplified magnetic field.

 Using the estimate of the obtained magnetic field, we can estimate the magnetic field of the Cas A
 progenitor $B_p=B_0R_fu_w/\Omega R^2_p$ where $\Omega $ is the rate of stellar rotation and
 $R_p$ is the radius of the progenitor. For characteristic values
 $R_p=10^{13}$ cm and $\Omega ={10^{-8}}$ s$^{-1}$ we obtain $B_p\sim 0.2$~G. 
 This number is significantly below than the value $B_p\sim 0.1-1$ kG inferred  from observations 
 of H$_2$O masers at distances $\sim 10^{15}$ cm from several RSGs with powerful stellar winds 
 (Vlemmings et al. \citealp{vlemmings05}). However, it is not excluded that the strong magnetic fields of RSG progenitor 
 dissipate at large distances in the stellar wind e.g. via magnetic reconnection. 

We can use the expression similar to Eq. (8)
also for the reverse shock. Substituting the reverse shock speed in the ejecta
frame, $V_{be}\sim 4.2\cdot 10^3$ km s$^{-1}$,  and  $M_A^b=8$ we
find the magnetic field in  ejecta
 $B_{ej}\sim 0.3\mu $G.
%For simplicity we assumed equal values of $p_0$ at the
%forward and reverse shocks.

 The maximum energy of protons $\sim $ 40 TeV in the  hadronic model H1
 is in agreement with previous
 estimates of maximum energies attainable at fast shocks with non-resonant streaming instability
(Zirakashvili \& Ptuskin \citealp{zirakashvili08}, Bell et al. \citealp{bell13}).
%However in this case the corresponding lower maximum energies of the protons accelerated
%are below the "knee" energy $\sim 3$ PeV in the observable cosmic ray spectrum.

Probably the obtained  low value of the strength of the magnetic field in the stellar wind
implies  that the stellar wind is not a
 good place for acceleration of cosmic rays up to PeV energies.
 If the forward shock of Cas A would propagate  in the
 interstellar medium with the magnetic field strength $\sim 5$ $\mu $G,
 the current value of the maximum energy
would be a factor of $\sqrt{400}$ higher, namely  800 TeV.

Generally, the high-energy part of the observed  spectrum is well reproduced in the leptonic model L1
(see Figure 8). However the low energy part of the spectrum
below 1 GeV  contributed by   non-thermal  bremsstrahlung
 of electrons, does not agree  with {\it Fermi} LAT  data.

To avoid  overproduction of the non-thermal  bremsstrahlung,   in the hadronic model H1 we assume
that  the magnetic field at the reverse shock is rather large.
It is interesting that this is possible only due to the presence of the RSG shell. As
mentioned in  Section 8.1,  this shell results in an  effective  transfer of the explosion energy in the
reverse shock region and in a stronger magnetic field. Because of the lack of such a shell
in the hadronic model H2,  the energetics of the reverse shock is
low, and the magnetic field is rather  weak. Therefore,  the number of relativistic
 electrons  must be very high  to reproduce the
 radio flux observed from  the reverse shock. Then,  the non-thermal  bremsstrahlung produced by
 same electrons might significantly exceed the gamma-ray flux  observed
at  sub-GeV energies  (see Figure 7).

\subsection{Amplification of magnetic field and maximum energies of accelerated particles}

The hadronic models demand  strong amplification of the  magnetic field
in order to reproduce the shape of the observed  broad-band energy
spectra of gamma-rays. For the
model H1,  the required  upstream values are
0.4 mG and 0.2 mG  for the forward and reverse shocks, respectively.
The Alfv\'en drift
in such fields may lead to steepening of the energy spectra of accelerated
particles.  This effect  has been earlier  recognized and explored
by Ellison et al. (\citealp{ellison99}) for Cas~A, and recently also by Morlino and
Caprioli (\citealp{morlino12}) for Tycho.

The comparison of the magnetic fields in  the stellar wind and ejecta
implies  very large  amplification factors: $3\cdot 10^4$ and 700 for the forward and reverse shocks,  respectively.
The magnetic field of the ejecta is the field of the exploded supernova progenitor.
Presumably,  the field was additionally amplified
 during the propagation of the radiative shock inside
the exploded star when vortex motions downstream of the shock
result in the efficient gas mixing (see the modeling of Iwamoto et
al. \citealp{iwamoto97}). These motions can also amplify random
magnetic fields in the star interior. The
random magnetic field may be amplified downstream of the shock
moving in the medium with density perturbations (Giacalone \&
Jokipii \citealp{giacalone07}). Eventually the shock crosses the
outer layers of the exploded star and  enters into a circumstellar
medium. The downstream material forms a supernova ejecta.

In order to estimate the ejecta random magnetic field  in this scenario,
let us assume that a small fraction $\xi _B$
of a supernova mechanical energy $E_{SN}$ is transformed into a magnetic energy. At later epochs
the magnetic field strength drops proportional to $R^2_p/V^2_{ej}t^2$, where $R_p$ is the
radius of a progenitor and $V_{ej}\sim (2E_{SN}/M_{ej})^{1/2}$ is the characteristic velocity of
ejecta. Hence the magnetic field of ejecta at time $t$ is determined as

\[
B_{ej}=\frac {M_{ej}}{t^2}\sqrt{\frac {3\xi _BR_p}{2E_{SN}}}=
\]
\begin{equation}
0.25\mu \mathrm{G}\xi _B^{1/2}\frac {M_{ej}}{M_{\odot}}
\left( \frac {E_{SN}}{10^{51}\mathrm{erg}}\right) ^{-1/2}
\left( \frac {R_p}{10^{13}\mathrm{cm}}\right) ^{1/2}
t^{-2}_{\mathrm{kyr}}.
\end{equation}
Even for the  modest value of $\xi _B=0.01$ the magnetic field in the ejecta of a
core collapse supernovae at $t<10^3$ yr achieves $\mu G$ level.
Lower values $B_{ej}\sim 10^{-9}-10^{-8}$G are expected for Ia supernovae originating
from a thermonuclear explosion of a white dwarf with a characteristic radius $R_p\sim 10^9$ cm.

In the case of Cas~A,  we obtain the magnetic field of ejecta $\sim $ 0.5 $\mu $G which is close to the
value found in the previous Section. Although this field is random,  it has  a rather large scales,
namely an
order of 1/100 of the radius of the reverse shock. It can be
 considered as a  large scale field because the non-resonant streaming instability amplifies magnetic fields with even smaller scales.

We should note that the above amplification factors are significantly larger
compared to the results obtained from the numerical modeling  of nonresonant
 streaming instability (Zirakashvili \& Ptuskin \citealp{zirakashvili08}). To a large extent
 this is not a surprise;
 it is difficult to amplify the magnetic field  by three orders of magnitude
 in MHD modeling with its limited spatial
 resolution.  It is likely that  the amplified magnetic fields are strongly underestimated in
the available numerical simulations because of the numerical viscosity.

Another possibility can be  related to a possible
additional amplification that takes place in the upstream regions of CR modified shocks moving in the medium
 with strong density perturbations. The deceleration of gas clumps
 by the CR pressure is slower  compared to  a less dense plasma.
 They will stretch and additionally amplify the random magnetic
 field pre-amplified far upstream of the shock by the non-resonant streaming instability.
The possibility of this kind of
 magnetic amplification was demonstrated recently by  Drury and Downes (\citealp{drury12}).
A similar  idea was also proposed earlier  by Beresnyak et al.
(\citealp{beresnyak09}). Given  that the stellar winds and ejecta
are very clumpy, so this scenario of additional amplification
seems quite plausible.
%This additional amplification might help to resolve another problem related with a small-scale nature of
%the non-resonant instability. If this amplification occur then the gyroradius  of highest energy particles will
%drop down closer to the scale of the generated random magnetic field.
%The diffusion of particles will be closer
%to the Bohm limit. Probably this will not result in the significant increase of
%the maximum energy of accelerated particles. This is because the shock modification at moderately
% modified shocks is mainly produced by low energy particles. Then the additional magnetic amplification will
% take place only in the narrow region close to the shock front where low energy particles are accelerated.
% Then the additional magnetic amplification will not strongly influence on the acceleration of highest
%energy  particles.

It should be noted that the high magnetic fields found in our modeling are not rare in young supernova remnants. For example Chandra et al. (\citealp{chandra04}), based on  the observed effect of   synchrotron aging of radio electrons in the  extragalactic IIb SNR 1993J,  derived the downstream magnetic field strength between 0.19~G and 0.33~G after 3200 days of the explosion. This value is also in agreement with the  synchrotron self-absorption model of  Fransson and Bj\"ornsson (\citealp{fransson98}). For comparison,
in our hadronic model H2 for Cas~A the downstream magnetic field strength is only 0.04 G at the same
3200th day  after the explosion.

The upper limit of the maximum energy of accelerated protons at the forward shock is estimated as (see
 Zirakashvili \& Ptuskin \citealp{zirakashvili08})

\[
E_m=21\ \mbox{TeV}\
n_H^{1/2}m_{\exp }R_{f,pc}
 \ln ^{-1}\left( B_0/B_b\right) \times
\]
\begin{equation}
\left( \frac {\eta _{esc}}{0.05}\right)
\left( \frac {V_f}{10^3\
\mbox{km\ s}^{-1}}\right) ^2.
\end{equation}

Here $\eta _{esc}=2F_E/\rho _0V_f^3$ is the ratio of the energy
flux of run-away particles $F_E$ to the flux of kinetic energy
$\rho _0V_f^3/2$, $B_b$ is the strength of the random magnetic field
in the circumstellar medium.

This limit corresponds to the situation when the random magnetic field is amplified via the
nonresonant streaming instability exponentially in time 
(see also Bell et al. \citealp{bell13}, Schure \& Bell \citealp{schure13}). The instability
 is driven by the electric current of the highest energy particles.

For estimates, below  we  use the value of logarithm $\sim 10$ in the denominator
of Eq. (10), and fix  the expansion parameter $m_{\exp }=0.7$.
The parameter $\eta _{esc}$ is determined by the energy density of particles $\epsilon $
at the end of their spectrum,  just before the cut-off.
For shocks with compression ratio $\sim 4$ the parameter $\eta _{esc}$ is estimated as
$\eta _{esc}\sim 0.5\epsilon/\rho _0V_f^2$. Using Fig.3,
we find $\eta _{esc}\sim 0.01$. Then one obtains
$E_m\sim 14$ TeV for the forward shock. For the  reverse shock
 we can use an  expression similar to Eq. (10). Using Fig. 3 we estimate
 $\eta _{esc}=0.05$ and obtain $E_m\sim 30$ TeV for the reverse shock.

From comparison of  these numbers with Fig.3 we see that the maximum
energies found in the modeling are significantly  larger:
$\sim 40$~TeV and $\sim 80$~TeV for
the forward and reverse shocks, respectively (the maximum energy of
oxygen ions 40 TeV per nucleon corresponds to the maximum energy
80 TeV for protons).

Since the parameters of our modeling are  adjusted to reproduce
the multiwavelength observations of Cas~A,  we conclude that both magnetic amplification and
maximum energy of accelerated particles are underestimated by the
theory, despite the  rather modest maximum energy of  about 40~TeV
required to explain the spectral points at very high energy gamma-rays.

The reason is the steep energy distribution  of accelerated particles derived from
the gamma-ray data. This results in a  small value of $\eta _{esc}=0.01$ at the forward shock.
The steep particle spectrum itself
is  explained by the strong magnetic field and
correspondingly  by  the high Alfv\'en speed upstream the shock. This
result found  for Cas A  has a more general implication;
it shows that the strong  magnetic field does
not necessarily  provide  higher maximum energy of accelerated
particles.

We should note that in the more realistic case of oblique shocks the maximum energy is higher 
 than the one given by Eq. 10. This equation was derived for quasiparallel shocks. 
 Taking this into account we can conclude that maximum energy 
 of particles accelerated  in Cas A corresponds to the number density of
 the highest energy particles (see Eq.10). Both quantities are in agreement with avalable TeV gamma-ray data.

Apparently, despite  certain achievements in the treatment of  amplification of the
magnetic field in young SNRs in general,  further development of theory is needed for
better understanding this complex phenomenon regarding, in particular, the demand of
very large field  amplification by a factor of $10^3$.

\subsection{Electron injection}
The problem of electron injection remains  one of the poorly understood aspects of
the theory of    diffusive shock acceleration of particles in SNRs. This general
problem becomes especially important in the models of Cas A without a RSG shell
(models H2 and L1) which require  very high electron-to-nucleon ratio  $K^b_{en}>0.1$ at the
 reverse shock. This  exceeds significantly the ratio
 $\sim 0.01$ observed in CRs.
 For this reason we invoke the electron (positron) injection from $^{44}$Ti decays
(Zirakashvili \& Aharonian \citealp{zirakashvili11}). To our knowledge, no other
ideas have been proposed so far  which would
 provide such a  high electron-to-nucleon ratio.

Note that  the electron injection from $^{44}$Ti decay is not required in
the model H1 in which $K^b_{en}=0.004$. Thus, any other electron
injection mechanism which can provide this
ratio at the reverse shock can be used in the model H1. However, the
positrons from $^{44}$Ti decays  accelerated later by the reverse shock
are of certain interest from the point of view of  origin of Galactic CR positrons.

The reason of high values of $K^b_{en}$ in the models H2 and L1
is the low energetics of the reverse shock. Correspondingly, the
amplified magnetic fields are weak. Thus, to reproduce the radio flux  from the reverse shock,
one has to assume large number of injected nonthermal electrons.

The forward shock of Cas A propagates in the stellar wind with a $r^{-2}$ density profile. In this environment the
energy of the shocked ejecta consists only 1$\% $ of the total energy in shocked plasma. The energetics of  reverse
 shocks is higher and is of the order of $10\% $ for  SNRs expanding  in the uniform medium. It can reach up to
 $50\% $ for SNRs expanding in the bubbles where plasma density increases towards the walls of the bubble. This is
 exactly what happened 40 years after explosion in the model H1. The collision of the shock  with the
 RSG shell transfers a significant fraction of the explosion energy into the region of the
 reverse shock.

 As for the electron injection at the forward shock of Cas A,
 we used a standard approach and prescribed some
 injection efficiency in models H1 and H2.
 Because of the low wind density, the amount of suprathermal  electrons
 produced via Compton scattering of gamma-rays from the decay of $^{56}$Co
 appears to be not sufficient  for  explanation of  the radio and
 X-ray production by the  forward shock of Cas A.
 At first glance  one may  increase the injection momentum of
 electrons to reproduce observations. However, any assumption on
 injection momenta higher than 200-300 MeV/c  leads to contradiction with the
 radio spectrum.

%On the other hand the radioactive material can be transported to the circumstellar medium by fast ejecta knots.
%It is known that FMKs are observed even beyond the shell of Cas A. Probably the less dense ejecta clumps
% are already destroyed and mixed with the circumstellar medium. The electron to nucleon ratios at the forward
% and reverse shocks differ a factor of 20 in the model H1. This means that if 1/20 fraction of the $^{44}$Ti
% reached the circumstellar medium then this will be enough for the explanation of electron injection at the
% forward shock.
%
%The interaction of the multiple fast ejecta knots with circumstellar medium results in
% its heating by bow shocks. This must result in the X-ray halo around the SNR. Such a halo is indeed observed but
% is attributed to the scattering of X-rays by dust.
%
% However the observation of the X-ray halo from the shock heated
% ambient medium of Cas A was reported recently by Lee et al. \cite{lee13}. They explain the observations
%  by  the propagation of  the forward shock far beyond the observed outer X-ray filaments. The X-ray
% filaments are related with the  forward shock eighther but they are the parts of the forward shock where the
% particle acceleration is effective.
%
% Nevertheless the bow shock interpretation of this result seems also possible.

%\section{Injection of hadronic component}

%The main difference between hadronic models H1, H2 and leptonic model L1  is the injection efficiency at the
% forward shock.

\subsection{On the need of acceleration at the reverse shock}

While the acceleration of particles by the reverse shock in Cas~A seems to be a
natural and well justified component in the general picture, it is fair to ask
a question whether
can we explain the multiwavelength data of Cas~A without
invoking the contribution of the reverse shock.  Formally,  the available data
do not  allow us to give a certain answer to this question.

Note that a nonlinear DSA model without the consideration of the reverse shock
has been discussed  by Berezhko et al. \cite{berezhko03}.
Their model contains many components of our models H1 and H2,
including the presence of the RSG shell.

In the model H2 almost all non-thermal X-rays are produced at the
forward shock while the thermal bremsstrahlung from the reverse
shock  contributes to X-rays at energies below
 10 keV. The production of  non-thermal X-ray emission at the reverse shock is
 a favored option  (Uchiyama \& Aharonian \citealp{uchiyama08}, Helder \& Vink \citealp{helder08}),
however it is not yet firmly established. In particular, it has been argued
that the non-thermal X-ray filaments in the reverse shock region are in fact
the projected filaments and belong to the forward shock (Delaney et al. \citealp{delaney04}).

The observed bright radio-ring can also be explained without the
electron  acceleration   at the reverse shock. The development of the
 Reighley-Taylor instability at the contact discontinuity can result in an additional
 amplification of magnetic field in the SNR interior (see e.g. the MHD modeling of Jun and Norman \citealp{jun96}), thus  the electrons accelerated at the forward shock may produce
 enhanced synchrotron radio-emission inside the shell of SNR.
% {\bf [it is too pessimistic regarding the impact of the reverse shock?]}

So, perhaps one  may argue that the major  morphological and spectral features
of Cas~A reported so far, can be addressed, in principle, by the forward shock.
On the other hand one may ask another question about the
fundamental difference between the reverse and forward  shocks in Cas A.
Both shocks  propagate in the media  having a
common origin. They are  the  same stellar material ejected prior to the explosion
 (the stellar wind) and during the explosion (the supernova ejecta).
 As argued  in Section 8.2,  presently the magnetic fields in the wind and ejecta are comparable.
 The speeds of both shocks also are similar.  Therefore,  there is no reason for DSA to operate effectively in the forward shock and  be suppressed in the  reverse shock.
The situation could  be  quite different for SNRs produced in Ia supernova
explosions. The magnetic  field of ejecta in these SNRs is
significantly lower  compared to  the interstellar magnetic
field (see Section 8.2).  Thus only the forward shock can contribute to the
very high energy radiation.
However,  even at such small fields, the reverse  shock can accelerate electrons
to GeV energies, and thus be a source of radio emission and gamma-ray bremsstrahlung.
Such a scenario  presumably is  realized in the Tycho where
non-thermal X-ray emission is not observed in the reverse shock region while
we see radio-emission related with the supernova ejecta (Dickel et al. \citealp{dickel91}).

\subsection{On the origin of radio-knots}

One of the distinct features of the reverse shock is the high
electron-to-nucleon ratio at the reverse shock, $K^b_{en} \sim 0.004-0.8$. Since the
reverse shock is modified by the pressure of accelerated ions, a
significant part of energy goes into accelerated electrons at the
reverse shock. In this regard a plasma of the reverse shock have
properties similar to properties of bright radio-knots.
Atoyan et al. \cite{atoyan00} found that the  energy of relativistic
electrons in the knots is close to $10^{-3}$ fraction of the
explosion energy. Since a volume factor of the knots is of the
order of $10^{-2}$ it is easy to estimate that the energy density of
accelerated electrons in the radio-knots may be comparable with
the mean energy density inside the supernova shell. This may be
considered as an independent evidence for a common origin of
radio-knots and supernova ejecta. In our model the high energy
density is explained by the "radioactive" origin of electrons and
positrons injected at the reverse shock (Zirakashvili \& Aharonian
\citealp{zirakashvili11}).

In the inner region of Cas~A,  the sheet-like structures and filaments are observed
in the infrared band   (Isensee et al. \citealp{isensee10}).
This material has not yet encountered the reverse shock.
These clumpy structures
 can produce the X-ray, optical and radio-knots knots after the reverse shock passage.

It is expected that the ejecta clumps have a broad range of densities. The clumps with the lowest
 density contrast move with a  velocity close to the the gas velocity just downstream of the
 reverse shock. This velocity $u\sim 1600$ km s$^{-1}$ in the model H1 (see Fig.2).
 The corresponding expansion age $T=R_b/u\sim 10^3$ years is close to the measured value of
 the radio-knots of the bright radio-ring
 (Tuffs \citealp{tuffs86}, Anderson \& Rudnick \citealp{anderson95}). The large-scale expansion of the
 radio-ring is even slower and is treated as the expansion of the reverse shock with the speed $(1160\pm 500)$
 km s$^{-1}$  (Delaney \& Rudnick \citealp{delaney03})
 which is comparable with the reverse shock velocity $V_b=770$  km s$^{-1}$ in the model H1.

Denser ejecta clumps are less decelerated after the reverse shock crossing and are initially
 observed as X-ray knots,  while the densest clumps are initially observed as optical knots.
At later epochs they have
entered the region downstream the forward shock. The
radio-electrons accelerated during the transition through the
reverse shock are still inside the clumps. Alternatively they  can be
accelerated by internal shocks in the clump.
The clumps are
eventually destroyed by the shear Kelvin-Helmholz instability.
A material and accelerated electrons of clumps are mixed with the
plasma of radio-plateau (see discussion and modeling of Anderson et al. \citealp{anderson94}).
The electrons emit a radio synchrotron
radiation in the strong magnetic field downstream of the forward
shock. This field probably is additionally amplified by the shear
instability. Since the reverse shock is strongly modified by CR
pressure,  the electron spectra at the reverse shock and in the
clumps have concave shape and are steeper at low energies. This
results in steep radio-spectra of the reverse shock and  knots.
The electrons that escape  the knots may produce even steeper
radio-spectra because of higher magnetic fields and because of an
energy dependent propagation as discussed by Atoyan et al.
\cite{atoyan00}. In addition, the radio-knots near
the forward shock are observed with  steeper spectra (Anderson \& Rudnick
\citealp{anderson95}).  A possible interpretation could be that closer
to the forward shock the magnetic field is stronger.  Here, 
the electrons emitting at the given radio frequency have lower energies. Then 
their concave spectrum results in the steeper spectra of the knots near the forward shock.

%On the other hand the RSG wind where the forward shock propagates is probably also
% very clumpy (Chevalier \& Oshi \citealp{chevalier03}). The densest clumps are observed inside
% the SNR in optics
% as slow-moving quasi-stationary flocculi (QSF) (Kamper \& van der Bergh \citealp{kamper76}).

%On the other hand some part of ejecta clumps containing the radioactive material can
%reach the circumstellar medium (see Sect. 8.4). The corresponding suprathermal electron and positrons
% are picked up  by the forward shock and accelerated.
%So we do not exclude that two different kinds of clumps (ejecta clumps and the stellar wind material
%mixed with the radioactive material) will result in two
% different sub sets of the radio-knots.

This qualitative picture  seems in agreement both with theoretical (Berezhko et al. \citealp{berezhko03};
Atoyan et al. \citealp{atoyan00}) and observational
conclusions (Anderson \& Rudnick \citealp{anderson95,anderson96}; Delaney \& Rudnick \citealp{delaney03}).

\section{Summary}

Below we summarize the main results of  study of the present work
on the nonlinear diffusive shock acceleration in Cas~A, and the related
broad-band electromagnetic radiation.

1. The available observational properties of Cas A are better reproduced by the hadronic model H1
when both the forward and reverse shocks significantly contribute to the broad-band emission.
The gamma-ray  production is strongly dominated by  decays of $\pi^0$-mesons from
interactions of protons (in the forward shock) and ions (in the reverse shock).
The contributions of the
forward and reverse shocks around 1 TeV are comparable. The reverse shock produces
also approximately 50\% of the radio, and 90\% of the X-ray synchrotron emission of Cas A.

2. The model  requires  ejected mass of about $M_{ej}=2M_{\odot }$ and the explosion energy
$E_{SN}=1.2\cdot 10^{51}$erg. The forward shock propagates into  the stellar wind with the mass-loss rate
$\dot{M}\sim 2.2\cdot 10^{-5}M_{\odot }(u_w/20$ km s$^{-1}$)  yr$^{-1}$. Magnetic fields required by the model
significantly exceeds the strength of the magnetic field  allowed by the MHD treatment  of the nonresonant
streaming instability. Either the current MHD simulations
strongly underestimates the strength of the amplified magnetic field,   or an
additional amplification is needed. A possible mechanism could be
 related to  the clumpiness of ejecta and the circumstellar medium.

3. Another important ingredient of the model H1 is the presence of the dense shell of the compressed red 
supergiant stellar wind at the distance $R_s=1.5$ ~pc from the supernova progenitor. Contrary to the pure 
 stellar wind models H2 and L1,  this permits to explain several observational features of Cas A like the 
 fast decrease of the flux of non-thermal X-rays, the dominance of the reverse shock in the production 
 of non-thermal X-rays and the slow expansion rate of the bright radio-ring and radio-knots. 

4. The acceleration efficiency in Cas A should be very high. At present 25$\%$ of the energy of
 supernova explosion is transferred into accelerated particles.  The previous claims of
 the low acceleration efficiency  have been  based on the high gas density derived
 from X-ray observations. The plasma density  has been  strongly
 overestimated because the non-thermal
 X-ray component was not properly taken into account.

5. The reverse shock of Cas A is modified by the pressure of energetic ions, positrons and electrons.
The energy content in accelerated  electrons and positrons from radioactive decay of $^{44}$Ti
is  very close to $10^{48}$ erg or  0.1\%  of the explosion energy.

%5.  The available radio, X-ray and gamma-ray data can be explained in the hadronic and leptonic  models with CR acceleration by reverse and forward shocks. The forward shock produces the main part of gamma and hard X-ray emission. The soft X-ray and gamma-ray emission is produced mainly at the reverse shock via synchrotron, pion decay, IC scattering, thermal and nonthermal bremsstrahlung. The hadronic model H1 is in agreement with all available gamma-ray observations. However the maximum energy of the protons is of the order of 100 TeV in this model. The leptonic model L1 is free of this drawback but seems in contradiction with  the Fermi data at energies below 1GeV.

6. The maximum energy of accelerated particles in Cas A cannot exceed 100~TeV. This can be related to the weak
magnetic field of about $\sim 0.01\mu G$ in the red supergiant stellar wind where the forward shock propagates.
Because of this,  the scale of the random magnetic field  generated by the nonresonant streaming
 instability is  rather small  $\sim 10^{13}$ cm. The corresponding to the small-scale scattering
 $p^2$ type  dependence of the diffusion coefficient is favored by hard X-ray and TeV gamma-ray data.

7. More accurate spectroscopic measurements of the gamma-ray spectrum of  Cas A are highly desirable.
The measurements at sub GeV and multi-TeV energies will help to better distinguish between the hadronic and
leptonic  scenarios. In the case of hadronic
 origin of gamma-ray emission, the spectral shape at multi TeV energies will give an important information
about
 the maximum energies of accelerated particles.
It will be also interesting to search gamma-rays from  the
vicinity of
 Cas A. The dense gas shells formed by the winds of the Cas A progenitor seems to be  ideal targets for
production of gamma-rays by highest energy particles which already have left the remnant. 
 These shells, presumably
 situated at 5-30$'$ from the center of Cas A,  can be considered as  potential sources of 
 multi-TeV gamma-rays - presently or in the near future.  
 It depends on whether or not the high energy particles have already reached the shells. 

%The SNR Cas A is an excellent astrophysical laboratory for investigations in the different scientific
%directions like  plasma physics, nuclear physics, particle acceleration, cosmic ray origin, stellar
%evolution etc. The new theoretical and observational discoveries are coming.

\begin{acknowledgements}
VNZ acknowledge the hospitality of the Max-Planck-Institut f\"ur
Kernphysik, where the part of this work was carried out.
The work of VNZ was also supported by the RFBR grant in Troitsk.

\end{acknowledgements}

%\appendix

%\section{Fermi data}

%\figbox*{}{}{
%

\end{document}